\documentclass[a4paper]{article}
\usepackage[letterpaper,top=2cm,bottom=2cm,left=2.5cm,right=2.5cm,marginparwidth=1.75cm]{geometry}
\usepackage{amsmath,amssymb,array, natbib}
\usepackage{booktabs}
\usepackage{algorithm}
\usepackage{algpseudocode}
\usepackage{amsmath}
\usepackage{amssymb}
\usepackage{amsthm}
\newtheorem*{remark}{Remark}
\usepackage{subcaption}
\usepackage{multirow}
\usepackage{multirow}
\usepackage{makecell}
\usepackage{adjustbox}
\usepackage{appendix}
\usepackage{tikz}
\usetikzlibrary{positioning}
\usepackage{adjustbox} 
\usepackage{hyperref}

\newcommand{\cS}{\mathcal{S}}

\newcommand{\cT}{\mathcal{T}}

\newcommand{\cI}{\mathcal{I}}
\newcommand{\qmq}[1]{\quad\mbox{#1}\quad}

\newcommand{\ot}{\overline{t}}
\newcommand{\hatphi}{\widehat{\phi}}

\newcommand{\hatp}{\hat{p}}

\newtheorem{theorem}{Theorem}

\newtheorem{proposition}{Proposition}

\providecommand{\code}[1]{\texttt{#1}}
\providecommand{\pkg}[1]{\textbf{#1}}

\renewcommand{\appendix}{
 \setcounter{section}{0}%
  \setcounter{subsection}{0}%
  \renewcommand\thesection{\Alph{section}}
  \setcounter{equation}{0}
  \renewcommand{\theequation}{S.\arabic{equation}}
  \setcounter{figure}{0}
  \renewcommand\thefigure{S\arabic{figure}}
  \setcounter{table}{0}
  \renewcommand\thetable{S\arabic{table}}  
    \setcounter{lemma}{0}
  \renewcommand\thelemma{S\arabic{lemma}}  
    \setcounter{proposition}{0}
 
  }

\title{HNPclassifier: An R Package for Hierarchical Neyman-Pearson Classification}
\author{
  Lujia Yang\thanks{Hong Kong Baptist University, \texttt{25480847@life.hkbu.edu.hk}}
  \and
  Che Shen\thanks{City University of Hong Kong, \texttt{chshen3-c@my.cityu.edu.hk}}
  \and
  Shunan Yao\thanks{Hong Kong Baptist University, \texttt{yaoshunan@hkbu.edu.hk}}
  \and
  Lijia Wang\thanks{City University of Hong Kong, \texttt{lijiwang@cityu.edu.hk}}
}
\date{}    

\begin{document}

\maketitle

\begin{abstract}
    In multi-class classification problems, classes often have a natural priority ordering (e.g., cancer stages, COVID-19 severity levels, or air-quality categories). In such settings, it is important to prioritize correct identification of more severe classes and to control under-classification errors, which occur when an observation from a higher-priority class is misclassified into a lower-priority one. The Hierarchical Neyman–Pearson (H--NP) framework of \cite{wang2024hierarchical} was developed for ordered multi-class settings to prioritize under-classification error control; its H--NP umbrella algorithm provides high-probability control of under-classification errors at user-specified levels. This paper introduces the R package \href{https://cran.r-project.org/web/packages/HNPclassifier/index.html}{\pkg{HNPclassifier}}, which implements H--NP umbrella algorithms to construct H--NP classifiers using built-in learners such as logistic regression, random forests, and support vector machines, as well as user-supplied scoring functions, thereby enabling effective error control for ordered multi-class classification tasks.

\end{abstract}

\section{Introduction}

In statistics and machine learning, classification aims to predict class labels (categories) for new observations. In many applications, classes have different priorities and may follow a natural ordering. For example, in HIV screening \citep{meyer1987screening}, correctly identifying positive cases is typically prioritized over correctly identifying negative cases. In Alzheimer’s disease stage diagnosis \citep{xiong2006measuring}, disease stages form an ordered progression, and accurate identification of earlier stages is often more important for timely intervention than identifying later stages, with healthy controls being the lowest-priority group. Similarly, COVID-19 severity classification \citep{alballa2021machine} commonly uses ordered categories such as severe, mild, and healthy. In such settings, misclassifying a high-risk or severe case as a lower-risk or less severe category can be far more consequential than other mistakes. These errors are termed under-classification errors \citep{wang2024hierarchical}, and controlling them is a natural objective in priority-ordered classification problems.


Classical classification methods commonly maximize accuracy or minimize an overall classification error, often formulated as a weighted sum of class-conditional errors with weights typically taken to be the class proportions. However, in many applications these weights do not reflect the true class priorities. In particular, such objectives do not directly target control of the more consequential under-classification errors. To address this issue in binary classification, the \emph{Neyman--Pearson (NP) paradigm} \citep{cannon2002learning,scott2005comparison,tong2013plug} provides a principled approach: it controls the under-classification error (a.k.a., type I error in the binary setting) at a user-specified level while minimizing the remaining error (a.k.a., type II error).

Furthermore, to address more general multi-class classification settings, \citep{wang2024hierarchical,tian2025neyman} study error control in multi-class problems and extend the Neyman--Pearson (NP) framework beyond the binary case. In particular, \citep{wang2024hierarchical} focuses on priority-ordered classes and the resulting under-classification errors, proposing the \emph{Hierarchical Neyman--Pearson (H--NP) paradigm}. Suppose there are $\cI$ classes with class labels $[\cI] :=\{1, 2,\ldots, \cI\}$ ordered in decreasing importance. For $i\in[\cI-1]$, the $i$-th under-classification error is the probability of misclassifying an observation from class $j$ with $j > i$. The H--NP framework aims to control this 
$i$-th under-classification error below a user-specified level $\alpha_i \in (0, 1)$, while minimizing a weighted sum of the remaining classification errors.

Classification under the NP paradigm has been studied via both empirical risk minimization approaches \citep{casasent2003radial,scott2005comparison,han2008analysis} and plug-in approaches \citep{tong2013plug,zhao2016neyman}. Subsequently, the NP umbrella algorithm \citep{tong2018neyman} was developed to accommodate a wide range of scoring-based classifiers (e.g., logistic regression, random forests, and SVMs) while providing high probability Type I error control in binary classification. The umbrella algorithm is implemented in the widely used R package \href{https://cran.r-project.org/web/packages/nproc/index.html}{\pkg{nproc}}.
For the multi-class case, \citep{tian2025neyman} connects multi-class NP classification to cost-sensitive learning and proposes corresponding algorithms, implemented in the R package \href{https://cran.r-project.org/web/packages/npcs/index.html}{\pkg{npcs}}. However, this approach is not tailored to under-classification errors and does not provide the high-probability error control guarantees. Inspired by the binary NP umbrella algorithm, \cite{wang2024hierarchical} proposes an H--NP umbrella algorithm that achieves high-probability under-classification error control; specifically, for each $i \in [\cI]$, , it aims to ensure that the $i$-th under-classification error is at most 
$\alpha_i$
 with probability at least 
$1 - \delta_i$, where the violation rates $\delta_i$'s are also user-specified.

This paper introduces the R package \href{https://cran.r-project.org/web/packages/HNPclassifier/index.html}{\pkg{HNPclassifier}}, which implements H--NP umbrella algorithms for ordered multi-class classification with high-probability control of under-classification errors. To our knowledge, it is the first R package to provide high-probability asymmetric error control for multi-class classification. Moreover, we evaluate the algorithm’s performance via simulations and real-data examples in settings with $\cI > 3$  classes, extending prior empirical studies in \cite{wang2024hierarchical} that focus primarily on the three-class case. The rest of the paper is organized as follows. Section~\ref{sec:hnp_classification} defines under-classification error and introduces the H--NP paradigm, together with the design and construction of the H--NP umbrella algorithm and the resulting form of the H--NP classifier.  Section~\ref{sec:hnp_package} provides an overview of the \href{https://cran.r-project.org/web/packages/HNPclassifier/index.html}{\pkg{HNPclassifier}} package, including the main R functions and objects, along with example code illustrating the primary syntax and usage. Section~\ref{sec:simulation} reports simulation results assessing algorithmic performance. Section~\ref{sec:real
_data} presents two real-data analyses—air-quality classification and credit-risk analysis—demonstrating the broad applicability of the proposed methods. Finally, the paper concludes with a brief summary and computational details of the implemented tools.




\section{Hierarchical Neyman--Pearson (H--NP) classification}\label{sec:hnp_classification}

\subsection{An H--NP classifier}
The H--NP classification framework targets multi-class problems in which class labels have a natural priority ordering, and it is designed to control the most consequential errors, \emph{under-classification}, at user-specified levels. We consider a set of classes $[\mathcal{I}]$, where class $i$ is more important than $j$ if $i < j$. $(X, Y)$ denotes a random pair where $X \in \mathcal{X}\subset \mathbb{R}^p$ as feature vectors and $Y \in [\cI]$ as class label. A classifier $\phi: \mathcal{X}  \rightarrow [\cI]$ maps a feature vector $X$ to a predicted class label. Let  $P_i(\cdot)$ abbreviate the conditional probability $\mathbb{P}(\cdot|Y=i)$.  We define the following two types of errors:
\begin{align}
&\text{$i$-th under-classification error:} \quad R_{i\star}(\phi):= P_i\bigl(\phi(X) > i\bigr)\le \alpha_i, \quad \text{for } i  \in [\mathcal{I}-1]; \label{eq:under-error}\\
&\text{remaining error:} \quad R^c(\phi) := \mathbb{P}\bigl(\phi(X)\neq Y\bigr) - \sum_{i=1}^{\mathcal{I}-1} \pi_i R_{i\star}(\phi), \quad \text{where } \pi_i = \mathbb{P}(Y=i). \label{eq:def_of_error}
\end{align}

The H--NP framework aims to control the $i$-th under-classification errors  under the user-specified thresholds $\alpha_i \in (0,1)$ for $i \in [\cI - 1]$ at the population level. Subject to these under-classification error constraints, the framework then seeks to minimize the remaining misclassification error. The corresponding optimization problem can be written as
\begin{equation}\label{eq:hnp_goal}
\min_{\phi} \; R^c(\phi)
\quad \text{s.t.} \quad
R_{i\star}(\phi) \le \alpha_i, \;\; \forall i \in [\mathcal{I}-1].
\end{equation}

Under the H--NP framework, \cite{wang2024hierarchical} proposed a scoring-type H--NP classifier that uses $\mathcal{I}-1$ scoring functions, denoted by $T_1, T_2, \ldots, T_{\mathcal{I}-1} : \mathcal{X} \to \mathbb{R}$, together with a corresponding set of thresholds $(t_1, t_2, \ldots, t_{\mathcal{I}-1})$. Here, each scoring function assigns a score to an input observation for the purpose of determining its class label. Each pair $(T_i(X), t_i)$ is used to determine whether the observation should be assigned to class $i$ or deferred to the lower-priority classes $\{i+1, \ldots, \mathcal{I}\}$. This decision structure transforms the original multi-class classification task into a sequence of binary decisions. The resulting H--NP classifier is defined as
\begin{equation}\label{eq:np_classifier_form}
\begin{cases}
1, & \text{if } T_1(X)\ge t_1,\\
i, & \text{if } T_j(X)<t_j \text{ for all } j<i \text{ and } T_i(X)\ge t_i,\quad i=2,\ldots,\mathcal{I}-1,\\
\mathcal{I}, & \text{otherwise.}
\end{cases}
\end{equation}
Then, the $i$-th under-classification error takes the form
\begin{equation}
R_{i_\star}(\hat{\phi}) = P_i\left(\hat{\phi}(X) > i \right) 
= P_i\left(T_j(X) < t_j, \forall j \leq i \right),\label{transform}
\end{equation}
where $X$ is a new observation drawn.  We note that the construction of the scoring functions $T_i$ and the selection of the thresholds $t_i$ depend on the observed samples. Consequently, the resulting classifier $\hat{\phi}$ is random, whereas $R_{i\star}(\cdot)$ and $R^c(\cdot)$ are population-level quantities that depend on the underlying distribution of $(X,Y)$. As a result, it is generally difficult to control under-classification error almost surely. Instead, the H--NP umbrella algorithm outputs a classifier \(\hat{\phi}\) such that the population under-classification errors are controlled in the sense that
\begin{equation}\label{eq:high_prob_control}
    \mathbb{P}\bigl(R_{i\star}(\hat{\phi}) > \alpha_i\bigr) \le \delta_i, \qquad \forall i \in [\mathcal{I}-1],
\end{equation}
where the $\delta_i$'s are user-specified tolerance levels for the violation probabilities of the under-classification constraints. The following section introduces the details of the algorithm and explains how it achieves high-probability control of the under-classification errors.

\subsection{H--NP umbrella algorithm}

The H--NP umbrella algorithm relies on a sample-splitting strategy: some data subsets are used to train the scoring functions based on a chosen base classification method, while other subsets are used to select appropriate score thresholds so as to achieve high-probability error control. We denote the set of samples from class $i$ by $\mathcal{S}_i=\{X_j \mid Y_j = i,\, j \in [N_i]\}$, where $N_i$ is the sample size of class $i$. The algorithm randomly divides the samples from class $i$ into up to three subsets: $S_{is}$ ($i \in [\mathcal{I}]$) is used to train the scoring functions $T_i$, $S_{it}$ ($i \in [\mathcal{I}-1]$) is used to select the thresholds $t_i$, and $S_{ie}$ ($i = 2, \ldots, \cI$)  is used to evaluate the empirical remaining error.

To train the scoring functions, the algorithm use the combined training set \(\mathcal{S}_s = \bigcup_{i \in [\mathcal{I}]} \mathcal{S}_{is}\) together with a probabilistic base classifier (e.g., logistic regression, random forest, or support vector machines) to estimate the posterior probabilities \(\widehat{\mathbb{P}}(Y = i \mid X)\) for each class \(i \in [\mathcal{I}]\). The   scoring functions used in \cite{wang2024hierarchical} are then constructed based on these estimated posterior probabilities $T_1(X) = \widehat{\mathbb{P}}(Y = 1 \mid X)$ and $T_i(X) = \widehat{\mathbb{P}}(Y = i \mid X) / \sum_{j = i+1}^{\mathcal{I}} \widehat{\mathbb{P}}(Y = j \mid X)$ for $i = 2, \ldots, \mathcal{I} - 1$. These scoring functions are used as the default option in our package, although users may also specify their own scoring functions. We emphasize that, regardless of the form of the scoring functions, the threshold-selection step in the H--NP algorithm still guarantees high-probability control of the under-classification errors.

Given the scoring functions, the H--NP algorithm sequentially selects the thresholds \(t_1, \ldots, t_{\mathcal{I}-1}\) using the subdatasets \(\mathcal{S}_{it}\)'s and an order-statistics-based approach to achieve high-probability control. We first consider the selection of \(t_1\), which is established by the following general proposition.
\begin{proposition}\label{prop:t1}[Adapted from \cite{tong2018neyman,wang2024hierarchical}]
For any integers \(0 < k < n\) and any \(\alpha \in (0,1)\), define
\begin{equation*}
    v(k,n,\alpha) := \sum_{j=0}^{k-1} \binom{n}{j}\alpha^j(1-\alpha)^{n-j}.
\end{equation*}
For any $i\in[\cI]$, 
denote $\cT_i = \{ T_i(X) \mid X\in \cS_{it}\}$, and let $t_{i(k)} $ be the corresponding $k$-th order statistic. Further denote the cardinality of $\cT_i$ as $n_i$. Assuming that the data used to train the scoring functions and the left-out data are independent and $n_i \geq \log \delta_i/ \log (1 - \alpha_i)$, then given a control level $\alpha_i$, for another independent observation $X$ from class $i$,
\begin{equation*}
\mathbb{P}\left(P_i\left[T_i(X) < t_i \mid t_{i(k_i)} \right] > \alpha_i\right) \leq  \delta_i \,, \quad \forall  t_i \leq t_{i(k_i)}, \quad \text{where} \quad k_i = \max \{k \mid v(k,n_i,\alpha_i) \leq \delta_i\} \,.
\end{equation*}
\end{proposition}

Proposition~\ref{prop:t1} implies that, by naively selecting \(t_i \le t_{i(k_i)}\) for all \(i \in [\mathcal{I}-1]\), one can achieve the desired high-probability control in Eq~\eqref{eq:high_prob_control}. However, the bound \(t_{i(k_i)}\) is generally overly conservative because it does not account for the effect of the previously selected thresholds \(t_j\), \(j<i\), on the \(i\)-th under-classification error. Consequently, this choice may yield a large value of \(R^c(\hat{\phi})\). To address this issue, we derive less conservative upper bounds for the \(i\)-th threshold for \(i>1\) by conditioning on the previously selected thresholds. This leads naturally to a sequential threshold-selection strategy.  The following result establishes the effectiveness of this thresholding strategy.

\begin{theorem}\label{prop:ti}[\cite{wang2024hierarchical}]
Given the previous thresholds $t_1, \ldots, t_{i - 1}$, we define a subset of these scores depending on the previous thresholds as $\cT'_i = \{ T_i(X) \mid X \in \cS_{it},  T_1(X)< t_1, \ldots, T_{i - 1}(X)< t_{i - 1} \}$. We use $t_{i(k)}$ and  $t'_{i(k)}$ to denote the $k$-th order statistic of $\cT_i$ (defined in Proposition~\ref{prop:t1}) and $\cT'_i$, respectively. Let  $n_i$ and $n_i'$ be the cardinality of $\cT_i$ and $\cT'_i$, respectively, and $\alpha_i$ and $\delta_i$ be the prespecified control level and violation tolerance for the $i$-th {under-classification} error.
Assuming $n_i \geq \log \delta_i/ \log (1 - \alpha_i)$,  we set 
\begin{equation*}\label{eq:adjustment}
\hatp_i = \frac{n_i'}{n_i}\,,\,  p_i =  \hatp_i + c(n_i)\,,\, \alpha_i' = \frac{\alpha_i}{p_i}\,,\, \delta_i' = \delta_i - \exp\{- 2n_i c^2(n_i)\}\,,  
\end{equation*}
where $c(n) = \mathcal{O}(1/\sqrt{n})$. Let
\begin{equation*}\label{eq:threshold_2}
\ot_i = \begin{cases}  t'_{i(k'_i)}\,, & \mbox{if } n_i' \geq \log \delta_i'/ \log (1 - \alpha_i') \qmq{and} \alpha_i' < 1\,; \\
t_{i(k_i)}\,, &\mbox{otherwise}\,,
\end{cases}
\end{equation*}
where $k_i = \max \{k\in[n_i] \mid v(k,n_i,\alpha_i) \leq \delta_i\} \qmq{and} 
    k'_i = \max \{k\in[n_i'] \mid v(k,n'_i,\alpha'_i)\leq \delta_i'\}\,.$ Then,

\begin{equation*}\label{eq:control_2}
\mathbb{P}(R_{i\star}( \hatphi)  > \alpha_i) =\mathbb{P}\left( P_i\left[T_{1} (X)< t_1, \ldots T_i (X)< t_i \mid \ot_i \right] > \alpha_i \right) \leq \delta_i \qmq{for all} t_i \leq \ot_i\,.    
\end{equation*}
\end{theorem}
The upper bound \(\ot_i\) guarantees the required high-probability control of the \(i\)-th under-classification error while providing a tighter bound than that in Proposition~\ref{prop:t1}. The computation of the upper bounds \(\ot_i\) is summarized in Algorithm~\ref{alg:upper_bound}.

\begin{algorithm}[h!]

\caption{\textsc{UpperBound}$(S_{it}, \alpha_i, \delta_i, (T_1, \dots, T_i), (t_1, \dots, t_{i-1}))$}\label{alg:upper_bound}
\label{algorithm_1}
\textbf{Input:} The left-out class-$i$ samples: $S_{it}$; level: $\alpha_i$; tolerance: $\delta_i$;  score functions: $(T_1, \dots, T_i)$; thresholds: $(t_1, \dots, t_{i-1})$ 
\begin{algorithmic}[1]
\State $n_i \gets |S_{it}|$
\State $\{t_{i(1)}, \dots, t_{i(n_i)}\} \gets \text{sort } \mathcal{I}_i = \{T_i(X) \mid X \in S_{it}\}$
\State $k_i =  \max \{k \mid v(k,n_i,\alpha_i) \leq \delta_i\}$
\State $\bar{t}_i \gets t_{i(k_i)}$
\If{$i > 1$}
    \State $\mathcal{I}'_i \gets \{t'_{i(1)}, \dots, t'_{i(n'_i)}\} = \text{sort}\{T_i(X) \mid X \in S_{it},\ T_1(X) < t_1, \dots, T_{i-1}(X) < t_{i-1}\}$
    \State $\hat{p}_i \gets n'_i/n_i,\quad p_i \gets \hat{p}_i + c(n_i),\quad \alpha'_i \gets \alpha_i/p_i,\quad \delta'_i \gets \delta_i - e^{-2n_ic^2(n_i)}$
    \If{$n'_i \geq \log \delta'_i / \log(1 - \alpha'_i)$ \textbf{and} $\alpha'_i < 1$}
        \State $k'_i =  \max \{k \mid v(k,n'_i,\alpha'_i) \leq \delta'_i\}$
        \State $\bar{t}_i \gets t'_{i(k'_i)}$
    \EndIf
\EndIf
\State \textbf{Output:} $\bar{t}_i$
\end{algorithmic}
\end{algorithm}

Given \(t_j\) for \(j < i \le \mathcal{I}-2\) and the corresponding upper bound \(\ot_i\), the choice of \(t_i\) influences the attainable value of \(R^c(\hat{\phi})\). This effect is generally non-monotone; see \citet{wang2024hierarchical} for discussion and examples. Hence, identifying the optimal \(t_i\) requires an explicit search. Moreover, conditional on the other thresholds, choosing \(t_{\mathcal{I}-1} = \ot_{\mathcal{I}-1}\) minimizes \(R^c(\hat{\phi})\), so \(t_{\mathcal{I}-1}\) does not require separate optimization. The remaining subset of the data, $\cS_e = \bigcup_{i=2}^{\cI} \cS_{ie}$, is used to compute the empirical counterpart of \(R^c(\hat{\phi})\). The final step of the algorithm is therefore to determine an optimal set of thresholds \((t_1, t_2, \ldots, t_{\mathcal{I}-1})\) that satisfies the upper-bound constraints while minimizing the remaining error. We carry this out via a grid search over candidate threshold combinations. The full H--NP umbrella algorithm is summarized in Algorithm~\ref{algorithm_2}, where details of the grid-search procedure are also provided.

\begin{algorithm}[h!]
\caption{\textsc{H--NP umbrella algorithm for } $\mathcal{I}$ classes}
\label{algorithm_2}
\textbf{Input:} Sample: $S=\bigcup_{i\in[\mathcal{I}]} S_i$; levels: $(\alpha_1,\ldots,\alpha_{\mathcal{I}-1})$;  tolerances: $(\delta_1,\ldots,\delta_{\mathcal{I}-1})$; grid set: $A_1,\ldots,A_{\mathcal{I}-2}$
\begin{algorithmic}[1]
\State $\hat{\pi}_i \gets |S_i|/|S|,\ \forall i\in[\mathcal{I}]$
\State $S_{1s},S_{1t} \gets \text{Random split } S_1$
\State $S_{is},S_{it},S_{ie} \gets \text{Random split } S_i,\ \forall i=2,\ldots,\mathcal{I}-1$
\State $S_{\mathcal{I}s},S_{\mathcal{I}e} \gets \text{Random split } S_{\mathcal{I}}$
\State $S_s \gets \bigcup_{i\in[\mathcal{I}]} S_{is}$
\State $T_1,\ldots,T_{\mathcal{I}-1} \gets \text{A base classification method}(S_s)$
\State $\bar{t}_1 \gets \textsc{UpperBound}(S_{1t},\alpha_1,\delta_1,(T_1),\text{NULL})$
\State $\tilde{R}^c \gets 1$
\For{$t_1 \in A_1 \cap (-\infty,\bar{t}_1]$}  
  \State $\bar{t}_2 \gets \textsc{UpperBound}(S_{2t},\alpha_2,\delta_2,(T_1,T_2),(t_1))$
  \For{$t_2 \in A_2 \cap (-\infty,\bar{t}_2]$}
    \State $\bar{t}_3 \gets \textsc{UpperBound}(S_{3t},\alpha_3,\delta_3,(T_1,T_2,T_3),(t_1,t_2))$
    \State $\cdots$
    \For{$t_{\mathcal{I}-2} \in A_{\mathcal{I}-2} \cap (-\infty,\bar{t}_{\mathcal{I}-2}]$}
      \State $\bar{t}_{\mathcal{I}-1} \gets \textsc{UpperBound}(S_{\mathcal{I}-1,t},\alpha_{\mathcal{I}-1},\delta_{\mathcal{I}-1},(T_1,\ldots,T_{\mathcal{I}-1}),(t_1,\ldots,t_{\mathcal{I}-2}))$
      \State $\hat{\phi} \gets \text{a classifier with respect to } t_1,\ldots,t_{\mathcal{I}-1}$
      \State $\tilde{R}^c_{\text{new}} \gets \sum_{i=2}^{\mathcal{I}} \frac{\hat{\pi}_i}{|S_{ie}|}\sum_{X\in S_{ie}} \mathbb{I}\{\hat{\phi}(X) < i\}$
      \If{$\tilde{R}^c_{\text{new}} < \tilde{R}^c$}
        \State $\tilde{R}^c \gets \tilde{R}^c_{\text{new}};\quad \hat{\phi}^* \gets \hat{\phi}$
      \EndIf
    \EndFor
    \State $\cdots$
  \EndFor
\EndFor
\State \textbf{Output:} $\hat{\phi}^*$
\end{algorithmic}

\vspace{0.5em}
\noindent\textbf{\textit{Remark:}} Each \(A_i\) is a user-specified grid for threshold search; see Steps 9, 11, and 14. In practice, if the user does not provide a grid, the default choice is \(A_i=\mathcal{T}_i\). If the user chooses to skip the grid search, the algorithm sets \(A_i=\{\bar{t}_i\}\), meaning that the threshold is fixed at \(t_i = \bar{t}_i\) in \(\hat{\phi}^*\).
\end{algorithm}



\section{Package overview}\label{sec:hnp_package}

\noindent
The \href{https://cran.r-project.org/web/packages/HNPclassifier/index.html}{\pkg{HNPclassifier}} package implements the H--NP umbrella algorithm~\ref{algorithm_2} for hierarchical Neyman--Pearson classification with ordered classes. Its main function constructs an H--NP classifier by adjusting several built-in base classification methods, including logistic, svm, and randomForest, or by using user-supplied scoring functions, according to user-specified control levels \(\alpha_i\)'s and violation rates \(\delta_i\)'s. The package also includes functions for evaluating the performance of H--NP classifiers, with particular attention to under-classification errors, which are not typically addressed in existing R packages. In this section, we provide an overview of the major functions in the package and explain their inputs. Illustrative examples are given in the next section.


\begin{table}[h]
\centering
\resizebox{\textwidth}{!}{
\renewcommand{\arraystretch}{1.2}
\begin{tabular}{p{0.23\linewidth} p{0.15\linewidth} p{0.62\linewidth}}
\hline
\textbf{Function} & \textbf{Output} & Description\\
\hline

\hline

\texttt{hnp\_umbrella()} 
& \textbf{function}
& Constructs an H--NP classifier using user-specified base classification methods, under-classification control levels, and violation rates.\\

\texttt{hnp\_summary()} 
& \textbf{list}
& Summarizes the prediction performance of a fitted classifier, including the confusion matrix, under-classification error rates, remaining error rates, and overall classification error rates. \\

\texttt{hnp\_boxplot()} 
& \textbf{plots + list}
& Displays boxplots of error metrics for comparing two classifiers across repeated experiments. \\

\hline
\end{tabular}
}
\caption{Core functions in the \href{https://cran.r-project.org/web/packages/HNPclassifier/index.html}{\pkg{HNPclassifier}} package and their output types.}
\label{tab:hnp_functions}
\end{table}

Table~\ref{tab:hnp_functions} summarizes the usage and output types of the main functions \texttt{hnp\_umbrella()}, \texttt{hnp\_summary()}, and \texttt{hnp\_boxplot()}. We next describe the key input arguments for these functions.

\medskip

\noindent For the main function
\begin{verbatim}
hnp_umbrella(X, Y, levels, tolerances, importance_order,
             method = "randomforest", pretrained_model = NULL, input_is_score = FALSE, 
             grid_search = TRUE, grid_set = NULL, max_grid = 15, max_combinations = 2000, 
             hnp_split = NULL)
\end{verbatim}
we divide the arguments into two categories: required user-specified inputs and optional inputs with default values.



\noindent The following inputs must be specified by the user:

\begin{itemize}
\item \texttt{X}: The input data in the form of a matrix or \texttt{data.frame}. If \texttt{input\_is\_score = FALSE}, \texttt{X} should contain feature variables. If \texttt{input\_is\_score = TRUE}, \texttt{X} should contain class-specific scores.

  \item \texttt{Y}: A vector of class labels for the observations in \texttt{X}. 

  \item \texttt{levels}: A numeric vector specifying the under-classification control levels \((\alpha_1, \ldots, \alpha_{\mathcal{I}-1})\).

  \item \texttt{tolerances}: A numeric vector specifying the violation tolerances \((\delta_1, \ldots, \delta_{\mathcal{I}-1})\).
  
\item \texttt{importance\_order}: A vector specifying the ordering of the possible class labels in \(Y\), from most important to least important. Based on this ordering, the original class labels are mapped to \(1, 2, \ldots, \mathcal{I}\) when constructing the H--NP classifier. The resulting classifier function still returns the original class labels in its predictions.

\end{itemize}

\noindent The following inputs have default values but may be modified according to the user's needs:

\begin{itemize}

\item \texttt{method}: A character string indicating the base classification method used to train the scoring function. Supported choices are \texttt{"randomforest"}, \texttt{"svm"}, and \texttt{"logistic"}.

\item \texttt{pretrained\_model}: User-provided scoring functions that take \(X\) as input and return scores, or fitted base classification models for which \texttt{predict(...)} can be used to compute estimated posterior probabilities for each class. Detailed examples of supported input formats are presented in the next section. If \texttt{pretrained\_model} is not \texttt{NULL}, the algorithm skips internal scoring function training and directly uses the provided one.
  

\item \texttt{input\_is\_score}: A logical flag indicating whether \texttt{X} contains feature variables or class-specific scores. If \texttt{TRUE}, the algorithm directly treats \texttt{X} as score input, skips internal scoring function training, and ignores \texttt{pretrained\_model}. It then outputs an H--NP classifier function that takes scores as input and returns class labels.



\item \texttt{grid\_search}: A logical flag indicating whether to perform grid search. If \texttt{TRUE}, the function performs a grid search to select the optimal threshold combination by minimizing the empirical remaining errors; see Steps 9, 11, and 14 of Algorithm~\ref{algorithm_2}. If \texttt{FALSE}, the function skips the grid search, computes the threshold upper bounds sequentially, and uses them directly as the thresholds in the output classifier.
  
\item \texttt{grid\_set}: A set of numeric values used as the grid sets \(A_i\)'s in Algorithm~\ref{algorithm_2} for grid search. If \texttt{grid\_set = NULL}, the algorithm uses the default choice \(A_i = \mathcal{T}_i\).

\item \texttt{max\_grid}: A numeric value controlling the maximum size of the grid sets. If \(\lvert A_i \cap (-\infty, \bar{t}_i] \rvert > \texttt{max\_grid}\), the algorithm randomly samples \texttt{max\_grid} values without replacement from \(A_i \cap (-\infty, \bar{t}_i]\) to form a subset for use in the grid search.

\item \texttt{max\_combinations}: A numeric value specifying the maximum number of threshold combinations considered when searching for the optimal classifier. To ensure that the total number of recursive iterations remains below \texttt{max\_combinations}, the algorithm resets \texttt{max\_grid} to
\(
\min\{\texttt{max\_grid}, \left\lfloor \texttt{max\_combinations}^{1/(\mathcal{I}-2)} \right\rfloor \},
\)
thereby controlling the number of threshold combinations explored.


\item \texttt{hnp\_split}: A list of length \(\mathcal{I}\) specifying the class-wise sample split. For example, with three classes, the user may specify
\texttt{list(c(train = 0.3, threshold = 0.7), c(train = 0.4, threshold = 0.55, error = 0.05), c(train = 0.95, error = 0.05))}. For class \(i = 2, \ldots, \cI - 1\), \texttt{hnp\_split[[i]]} should be a numeric vector specifying, in order, the proportions of the data used for score training, threshold selection, and empirical evaluation of the remaining errors. For class \(i = 1\), it should specify the proportions used for score training and threshold selection, whereas for class \(i = \cI\), it should specify the proportions used for score training and empirical objective evaluation. If \texttt{hnp\_split = NULL}, the function uses the internal default split described in Table~\ref{tab:hnp_split_default}.

    



\begin{table}[ht]
\centering
\begin{tabular}{l|l|ccc}
\hline

\hline
\multicolumn{5}{l}{\textbf{Internal scoring function training}} \\
\multicolumn{5}{l}{(\texttt{pretrained\_model = NULL} and \texttt{input\_is\_score = FALSE})} \\
\hline
\texttt{grid\_search} & Class & Train & Threshold & Error \\
\hline
\multirow{3}{*}{\texttt{TRUE}} 
& \(1\) & 0.5 & 0.5 & -- \\
& \(2, \ldots, \mathcal{I}-1\) & 0.45 & 0.5 & 0.05 \\
& \(\mathcal{I}\) & 0.95 & -- & 0.05 \\
\hline
\multirow{3}{*}{\texttt{FALSE}} 
& \(1\) & 0.5 & 0.5 & -- \\
& \(2, \ldots, \mathcal{I}-1\) & 0.5 & 0.5 & 0 \\
& \(\mathcal{I}\) & 1 & -- & 0 \\
\hline

\hline
\multicolumn{5}{l}{\textbf{User-provided base classification models, scoring functions, or scores}} \\
\multicolumn{5}{l}{(\texttt{pretrained\_model != NULL} or \texttt{input\_is\_score = TRUE})} \\
\hline
\texttt{grid\_search} & Class & Train & Threshold & Error \\
\hline
\multirow{3}{*}{\texttt{TRUE}} 
& \(1\) & 0 & 1 & -- \\
& \(2, \ldots, \mathcal{I}-1\) & 0 & 0.95 & 0.05 \\
& \(\mathcal{I}\) & 0 & -- & 1 \\
\hline
\multirow{3}{*}{\texttt{FALSE}} 
& \(1\) & 0 & 1 & -- \\
& \(2, \ldots, \mathcal{I}-1\) & 0 & 1 & 0 \\
& \(\mathcal{I}\) & 0 & -- & 1 \\
\hline
\end{tabular}
\caption{Default values of \texttt{hnp\_split} under different settings.}
\label{tab:hnp_split_default}
\end{table}

\end{itemize}

The following two functions are used to evaluate the performance of the classifiers. Below, we provide a brief description of their inputs.

\medskip
\noindent
The first function summarizes the performance of classifiers numerically.

\texttt{hnp\_summary(classifier, X, Y, importance\_order = NULL)},

\begin{itemize}
\item \texttt{classifier}: A classification function. It may be generated by \texttt{hnp\_umbrella()} or obtained from a general classification model trained using other packages.

\item \texttt{X}: A matrix or \texttt{data.frame} containing the feature variables. If \texttt{classifier} takes scores as input, then \texttt{X} should contain the corresponding score values. A detailed example is provided in the next section.

\item \texttt{Y}: A vector of true class labels corresponding to the observations in \texttt{X}.

\item \texttt{importance\_order}: An optional character vector specifying the class priority order, from highest to lowest. If\texttt{importance\_order = NULL},  the algorithm will atuomatics factor the $Y$ label,and take the order based on the unicode as the prioty order, but the warning will give to the user to suggest them to specify their designed order

\item \texttt{importance\_order}: An optional character vector specifying the class priority order from highest to lowest. If \texttt{importance\_order = NULL}, the algorithm will automatically convert the response variable \texttt{Y} to a factor and use the default order of its levels (determined by their Unicode ordering) as the priority order. In this case, a warning is given to encourage the user to specify the order.



\end{itemize}

\medskip

\noindent
The second function presents performance of classifiers in the format of boxplots.

\begin{verbatim}
hnp_boxplot(conf_1, conf_2, 
            levels = NULL, tolerances = NULL, 
            name_1 = "Classical", name_2 = "H-NP")
\end{verbatim}

\begin{itemize}

\item \texttt{conf\_1}, \texttt{conf\_2}: Lists of confusion matrices obtained from repeated experiments. Each element of a list is the confusion matrix from one run, used to compute the errors of interest.

\item \texttt{levels}: A numeric vector specifying the target under-classification control levels \((\alpha_1, \ldots, \alpha_{\mathcal{I}-1})\). If \texttt{NULL}, the corresponding control levels are not shown as black dashed lines in the boxplots of the under-classification errors.

\item \texttt{tolerances}: A numeric vector specifying the target violation tolerances \((\delta_1, \ldots, \delta_{\mathcal{I}-1})\). If \texttt{NULL}, the corresponding \(1-\delta_i\) quantiles of the \(i\)-th under-classification errors are not marked by red dots in the boxplots.

\item \texttt{name\_1}, \texttt{name\_2}: Character strings used as the displayed method names in the generated plots.

\end{itemize}

\section{Implementation details}\label{sec:implementation_details}

 This section describes the implementation details of the package and provides practical guidance for users. We present two examples to demonstrate how H--NP classifiers can be constructed: in the first, classical classification methods supported by the package, such as logistic regression, random forests, and support vector machines, are adapted to the H--NP framework; in the second, user-trained scoring functions are incorporated, including those derived from newer methods for which no mature package implementation is yet available. Specifically, Example 1 uses the built-in logistic regression base learner, whereas Example 2 employs a customized neural network as the base learner.

\subsection{Example 1}

We consider an ordered three-class Gaussian classification problem with feature dimension $d=4$. For each class, we generate $500$ training observations and $500$ testing observations. The class labels are $A$, $B$, and $C$, ordered from most severe to least severe. For each class $k \in \{A, B, C\}$, each coordinate of its mean vector is sampled independently from the uniform distribution on $[-1.5,1.5]$. The covariance matrix takes the form $\Sigma_k = \sigma_k^2 I_d$, where $\sigma_k^2 \sim \mathrm{Unif}(1,3)$.


\begin{verbatim}
library(MASS)
set.seed(123)
d <- 4
feats <- paste0("x", seq_len(d))
labels <- c("A", "B", "C")
n_each_train <- c(500, 500, 500)
n_each_test <- c(500, 500, 500)

means <- lapply(1:3, function(i) runif(d, -1.5, 1.5))
sigma2 <- runif(3, 1, 3)
Sigmas <- lapply(sigma2, function(s) diag(s, d))
\end{verbatim}
We generate the training and testing sets from three Gaussian classes.

\begin{verbatim}
X1 <- mvrnorm(n_each_train[1], mu = means[[1]], Sigma = Sigmas[[1]])
X2 <- mvrnorm(n_each_train[2], mu = means[[2]], Sigma = Sigmas[[2]])
X3 <- mvrnorm(n_each_train[3], mu = means[[3]], Sigma = Sigmas[[3]])
X1t <- mvrnorm(n_each_test[1], mu = means[[1]], Sigma = Sigmas[[1]])
X2t <- mvrnorm(n_each_test[2], mu = means[[2]], Sigma = Sigmas[[2]])
X3t <- mvrnorm(n_each_test[3], mu = means[[3]], Sigma = Sigmas[[3]])
Train <- data.frame(rbind(X1, X2, X3)); Test <- data.frame(rbind(T1t, T2t, T3t))
names(Train) <- feats; names(Test) <- feats
Train$y <- factor(c(rep("A", nrow(X1)), rep("B", nrow(X2)), rep("C", nrow(X3))),
                    levels = c("A", "B", "C"))
Test$y <- factor(c(rep("A", nrow(X1t)), rep("B", nrow(X2t)), rep("C", nrow(X3t))),
                    levels = c("A", "B", "C"))
\end{verbatim}
Then, we construct a dataframe in the following form
\begin{verbatim}
> head(Train, 2)
      x1         x2         x3        x4 y
1 -3.4672475 -0.4756288  0.8795137  3.114252 A
2 -0.9296232 -0.2072562  0.2830425 -1.501698 A
\end{verbatim}

Next, we fit an H--NP classifier using a multinomial logistic base learner. The under-classification control levels and violation rates are set to \((\alpha_1,\alpha_2)=(0.1,0.1)\) and \((\delta_1,\delta_2)=(0.1,0.1)\), respectively.
\begin{verbatim}
library(HNPclassifier)
clf_hnp <- hnp_umbrella(X = Train[, feats], Y = Train$y,               
                        levels = c(0.1, 0.1), tolerances = c(0.1, 0.1), 
                        importance_order = c("A", "B", "C"), 
                        method = "logistic", max_grid = 30)
\end{verbatim}
\begin{verbatim}
> print(clf_hnp)
function(newX, output_internal = FALSE){...}
\end{verbatim}
The ouput \texttt{clf\_hnp} is the fitted H--NP classifier returned by \texttt{hnp\_umbrella()}. Here, ``\texttt{...}'' denotes the details of the R function. The function \texttt{clf\_hnp} takes observations in the same format as the
training features and returns corresponding predicted labels. Here, we use the first five observations from the testing dataset \texttt{Test} as an example.
\begin{verbatim}
prediction <- clf_hnp(Test[, feats])
> head(prediction)
[1] "A" "B" "A" "A" "A" "A"
\end{verbatim}
The object \texttt{clf\_hnp} also stores several key attributes of the trained H--NP classifier. Users can extract them using \texttt{attr()} function.
\begin{verbatim}
> attr(clf_hnp, "label_mapping")
  A   B   C 
 "1" "2" "3" 
> attr(clf_hnp, "thresholds")
[1] 0.0418210 0.5698053
> attr(clf_hnp, "objective")
[1] 0.36
\end{verbatim}
The attribute
\texttt{attr(clf\_hnp, "label\_mapping")} returns the priority ordering of the original class labels in the H--NP classifier. Here, larger integers correspond to lower priority, and \texttt{"1"} denotes the highest-priority class. The second attribute\texttt{attr(clf\_hnp, "thresholds")} returns the threshold vector \((t_1, t_2)\) used in the H--NP classifier, as shown in \eqref{eq:np_classifier_form}. The last one \texttt{attr(clf\_hnp, "objective")} reports the empirical \(R^c(\phi)\), defined in \eqref{eq:def_of_error}, on the training sample, which is computed using the left-out sample \(S_{ie}\) for the evaluation step.

For a given classifer function, the function \texttt{hnp\_summary} is useful for evaluating the fitted classifier.
\begin{verbatim}
out_metrics <- hnp_summary(classifier = clf_hnp, X = Test[, feats], Y = Test$y)
\end{verbatim}
The object \texttt{out\_metrics} contains summary statistics for evaluating the out-of-sample performance of the H--NP classifier on the testing sample. Below, we list the most relevant performance metrics. Among these, the confusion matrix and overall classification accuracy are the most informative summaries of the classifier's out-of-sample performance.
\begin{verbatim}
> print(out_metrics$confusion_matrix)
    Predicted
True   A   B   C
   A 475  15  10
   B 216 241  43
   C 300  19 181
   
> print(out_metrics$overall_accuracy)
[1] 0.598
\end{verbatim}

Under the H--NP framework, the under-classification error defined in \eqref{eq:under-error} is the primary performance measure. The object \texttt{out\_metrics} also records the empirical under-classification errors.
\begin{verbatim}
> print(out_metrics$under_classification_error)
[1] 0.050 0.086 0.000
\end{verbatim}
This vector reports the under-classification error rates for the classes, ordered from highest priority to lowest priority. Since under-classification error is defined only for classes other than the lowest-priority class, the under-classification error for the last class is set to \(0\). Besides, users also have access to the remaining error, defined in \eqref{eq:def_of_error}, evaluated on the testing data.
\begin{verbatim}
> print(out_metrics$remaining_error)
[1] 0.3566667
\end{verbatim}
Furthermore, users can directly compare the predicted class label and the true label for each observation as follows
\begin{verbatim}
> head(out_metrics$predictions, 2)
  true_class predicted_class
1          A               A
2          A               B
\end{verbatim}

\subsection{Example 2}
This example illustrates how to construct the H--NP classifier by incorporating a pretrained classification model, scoring function, or user-provided scores via the R function \texttt{hnp\_umbrella}. We consider a three-class classification problem with classes \(\{A, B, C\}\). For each class, \(500\) observations are generated independently for each of the following purposes: scoring function training, threshold selection, and testing. These observations are sampled independently from the uniform distribution on a three-dimensional ball with radius \(1.5\), where each entry of the corresponding ball center vector is independently sampled from the uniform distribution on \([-2, 2]\). For the user’s convenience, our package provides the function \texttt{gen\_ball\_data()} for generating data from the aforementioned ball distribution with user-specified radius and center vectors, as follows:
\begin{verbatim}
library(HNPclassifier)
n <- 500; d <- 3
radii <- c(1.5, 1.5, 1.5); centers <- replicate(3, runif(d, -2, 2), simplify = FALSE)
Train     <- generate_ball_data(n, centers, radii) 
Threshold <- generate_ball_data(n, centers, radii)
Test      <- generate_ball_data(n, centers, radii)
> head(Train, 2)
        x1         x2         x3 y
1 1.310906 -0.5403522 -0.6266248 A
2 1.214084  0.9339565 -0.5367670 A
\end{verbatim}

We first construct a pretrained model using a single-hidden-layer softmax neural network with \(8\) hidden units, which estimates class probability scores. The package also provides a simple function \texttt{train\_nn\_and\_get\_scores()} for constructing such a model, although users are free to use functions from other packages to build their preferred models.
\begin{verbatim}
feats <- paste0("x", 1:3)
nn_model <- train_nn_and_get_scores(X = Train[, feats], Y = Train$y)
\end{verbatim}
The object \texttt{nn\_model\$model} is the fitted neural network model. The corresponding class probability scores can be obtained using the \texttt{predict()} function, as shown in the following example.
\begin{verbatim}
nn_score <- predict(nn_model$model, newdata = Train[, feats], type = "raw")  
> head(nn_score, 2)
             A            B            C
[1,] 1.0000000 4.837291e-11 5.380741e-09
[2,] 0.8655549 1.007973e-08 1.344451e-01
\end{verbatim}

Given the fitted model \texttt{pretrained\_model = nn\_model\$model}, the function \texttt{hnp\_umbrella()} skips the scoring function  training step and directly uses the pretrained model to construct the H--NP classifier. Here, we consider classes \(\{C, A, B\}\) in descending order of priority and we skip the grid search, while the under-classification control levels and corresponding violation rates are the same as in Example~1.

\begin{verbatim} 
clf_model <- hnp_umbrella(X = Threshold[, feats], Y = Threshold$y, 
                         levels = c(0.1,0.1), tolerances = c(0.1,0.1), 
                         importance_order = c("C", "A", "B"),
                         pretrained_model = nn_model$model,  grid_search = FALSE)

> attr(clf_model, "method") 
[1] "pretrained"
\end{verbatim}
Similar to the previous example, we can use \texttt{hnp\_summary()} together with the testing data set to evaluate the fitted H--NP classifier.
\begin{verbatim} 
out_model <- hnp_summary(classifier = clf_model, X = Test[, feats], Y = Test$Y)
> print(out_model$confusion_matrix)
    Predicted
True   C   A   B
   C 464  24  12
   A 129 339  32
   B  99  51 350
        
> print(out_model$under_classification_error)
[1] 0.072 0.064 0.000

> print(out_model$overall_accuracy)
[1] 0.7686667
\end{verbatim}
We note that the classes in this confusion matrix are arranged according to the user-specified importance order.

Alternatively, \texttt{hnp\_umbrella()} can use a pretrained scoring function or a score matrix instead of a pretrained model. This gives users greater flexibility by avoiding reliance on available R packages alone and makes it easier to incorporate models trained or scores computed using other tools. Below, we illustrate this approach by constructing a scoring function and score matrices and then using them as inputs to build H--NP classifiers. We first derive the scoring function from the previously trained model \texttt{nn\_model\$model} as follows:

\begin{verbatim}
score_fun <- function(X) {
  out <- predict(nn_model$model, newdata = as.data.frame(X), type = "raw")}
> head(score_fun(Train[, feats]), 2)
             A            B            C
[1,] 1.0000000 4.837291e-11 5.380741e-09
[2,] 0.8655549 1.007973e-08 1.344451e-01
\end{verbatim}
We then construct the H--NP classifier by supplying \texttt{score\_fun} through the \texttt{pretrained\_model} argument:
\begin{verbatim}
clf_function <- hnp_umbrella(X = Threshold[, feats], Y = Threshold$Y, ...,
                            pretrained_model = score_fun)
out_function <- hnp_summary(classifier = clf_function, X = Test[, feats], Y = Test$Y)
> print(out_function$confusion_matrix)
    Predicted
True   C   A   B
   C 464  24  12
   A 129 339  32
   B  99  51 350
\end{verbatim}
The arguments that are identical to those used in constructing \texttt{clf\_model} are omitted here and denoted by ``\texttt{...}".

We then use the score matrices to construct the H--NP classifier and evaluate its performance:
\begin{verbatim}
score_threshold <- score_fun(Threshold[, feats])
score_test <- score_fun(Test[, feats])
clf_score <- hnp_umbrella(X = score_threshold, Y = Threshold$Y, ..., input_is_score = TRUE)
out_score <- hnp_summary(classifier = clf_score, X = score_test, Y = Test$Y)
> print(out_score$confusion_matrix)
    Predicted
True   C   A   B
   C 464  24  12
   A 129 339  32
   B  99  51 350
\end{verbatim}
If the user uses score matrices to construct the H--NP classifier, then the resulting classifier function takes score vectors as input and returns predicted labels. Accordingly, when using \texttt{hnp\_summary()}, the new data must be supplied as a score matrix, e.g., \texttt{score\_test}, corresponding to the testing data set, rather than as the testing data set itself.

\begin{remark}
    Three confusion matrices presented in this example are identical, which indicates that \texttt{\upshape clf\_model}, \texttt{\upshape clf\_function}, and \texttt{\upshape clf\_score} are exactly equivalent classifiers. Thus, whether the user supplies a pretrained base classification model, a scoring function, or score matrices, our function produces the same H--NP classifier regardless of the input format. Users therefore need not be concerned that the choice of input format will influence the performance of the resulting classifier.
\end{remark}

To assess whether the H--NP umbrella algorithm achieves effective high-probability under-classification error control, a single measure of the under-classification error for one H--NP classifier is not sufficient. Instead, we repeat the experiment and examine how often H--NP classifiers trained on different samples violate the desired control levels. The function \texttt{hnp\_boxplot()} summarizes the distributions of the error rates across repeated experiments and facilitates comparison between the underlying base classification methods and the H--NP classification method. It also provides a convenient way to assess the effectiveness of the high-probability control, as illustrated in the following example.

We use the same setting as before and repeat the experiment 100 times. We construct two lists, \texttt{conf\_classical} and \texttt{conf\_hnp}, to store the confusion matrices from each run, corresponding to the base classical classification method and the H--NP classification method, respectively. The function \texttt{hnp\_boxplot()} then takes these two lists as input.
\begin{verbatim}
conf_classical <- list(); conf_hnp <- list()
alphas <- c(0.1, 0.1); deltas <- c(0.1, 0.1); class_order <- c("C", "A", "B")
for (i in seq_len(500)) {
    Train     <- generate_ball_data(n, centers, radii)
    Threshold <- generate_ball_data(n, centers, radii)
    Test      <- generate_ball_data(n*100, centers, radii)
    base_model <- train_nn_and_get_scores(X = Train[, feats], Y = Train$y)
    clf_hnp <- hnp_umbrella(X = Threshold[, feats], Y = Threshold$y,
                            levels = alphas, tolerances = deltas, importance_order =  class_order,
                            pretrained_model =  base_model$model, grid_search = FALSE)

    out_classical <- hnp_summary(classifier =  base_model$model, X = Test[, feats], 
                                  Y = Test$y, importance_order = class_order)
    out_hnp <- hnp_summary(classifier = clf_hnp, X = Test[, feats],
                                 Y = Test$y, importance_order = class_order)
        
    conf_classical[[i]] <- out_classical$confusion_matrix
    conf_hnp[[i]]  <- out_hnp$confusion_matrix }
\end{verbatim}





\begin{figure}[!ht]
\centering
\begin{subfigure}[b]{0.6\textwidth}
    \centering
    \includegraphics[width=\textwidth]{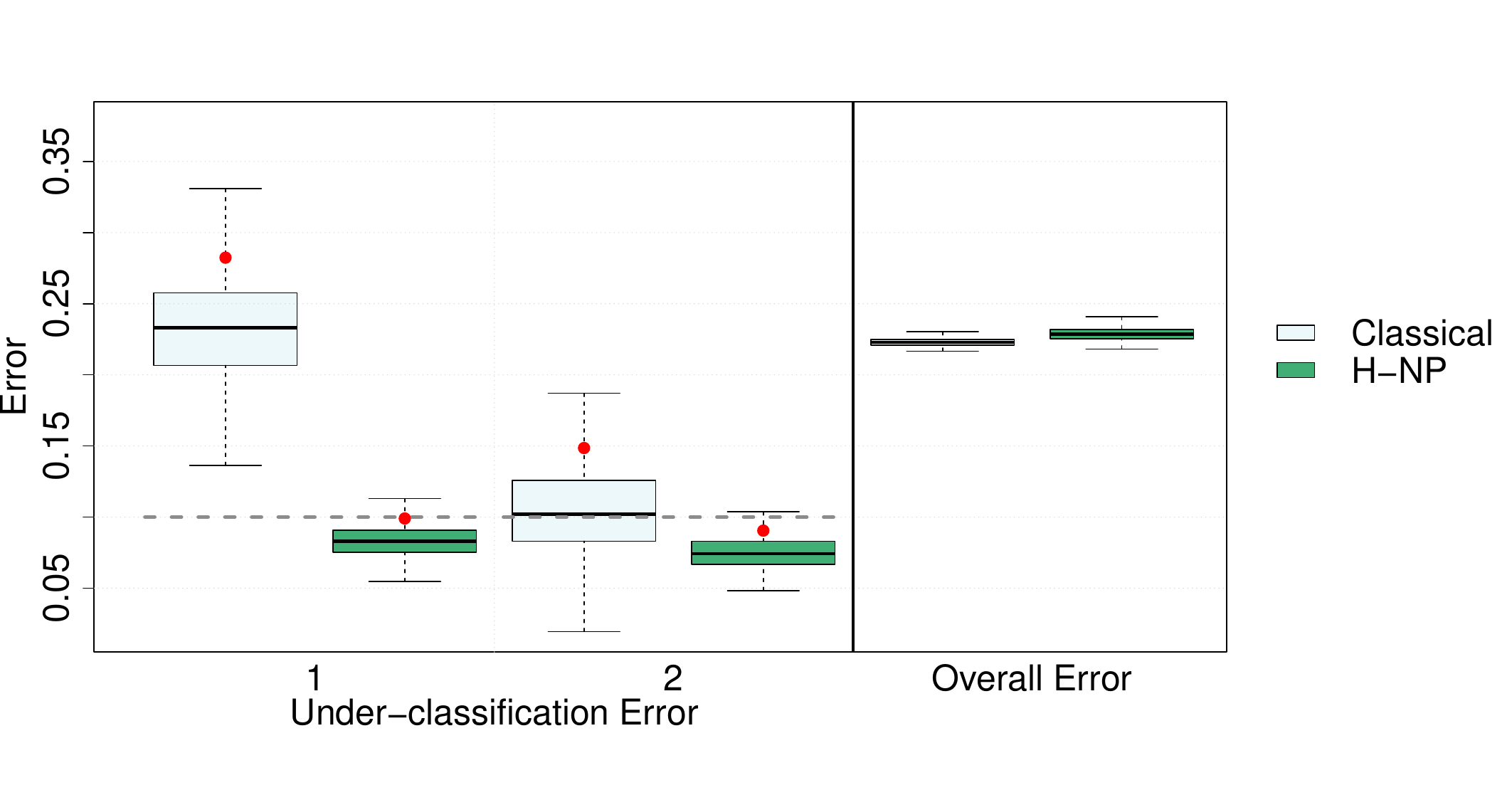}
          \vspace{-0.7cm}
    \caption{\(\alpha_1 = \alpha_2 = \delta_1 = \delta_2 = 0.1\)}
    \label{fig:example3_boxplots}
\end{subfigure}
\hfill
\begin{subfigure}[b]{0.6\textwidth}
    \centering
    \includegraphics[width=\textwidth]{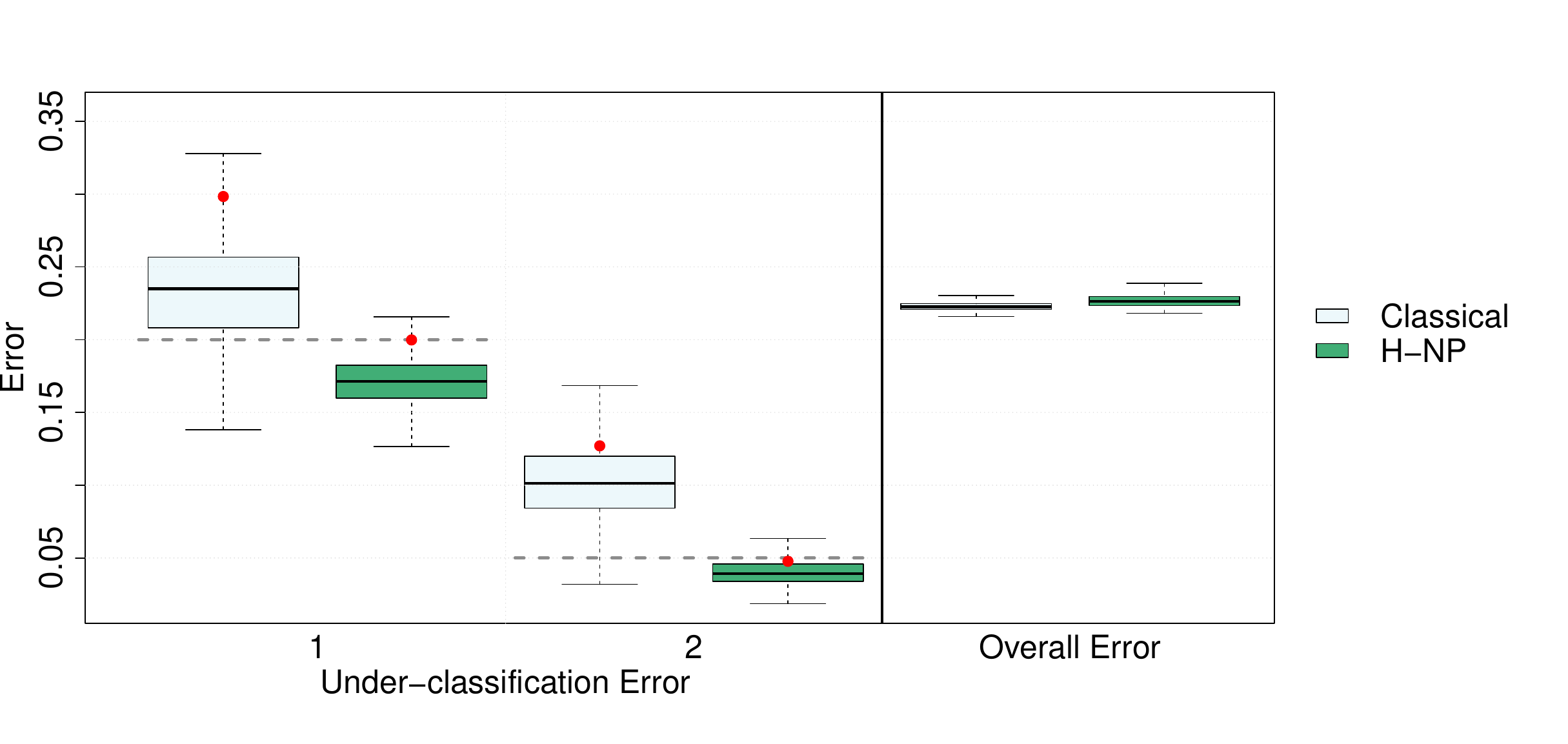}
    \vspace{-0.7cm}
    \caption{\(\alpha_1 = 0.2,\ \alpha_2 = 0.05,\ \delta_1 = 0.05,\ \delta_2 = 0.2\)}
    \label{fig:example3_boxplots_1}
\end{subfigure}
\caption{\footnotesize Distribution of the approximate errors for Example 2 under different control levels and tolerances.}
\end{figure}

Using the lists of confusion matrices, we can summarize the performance of the H--NP classification method and the corresponding base classical classification methods by calling the following code.
\begin{verbatim}
> hnp_boxplot(conf_1 = conf_classical, conf_2 = conf_hnp, levels = alphas, tolerances = deltas)
$classwise
                                Classical_Class_1 Classical_Class_2 H-NP_Class_1 H-NP_Class_2 
control level                          0.10000000        0.10000000    0.1000000   0.10000000
tolerance                              0.10000000        0.10000000    0.1000000   0.10000000
under-classification error mean        0.23128388        0.10431428    0.0836678   0.07501768
under-classification error sd          0.04170795        0.03197487    0.0114961   0.01137268
violation rate                         0.99600000        0.53000000    0.0840000   0.01400000

$overall
                                       Classical        H-NP
overall misclassification error mean 0.223030413 0.229063360
overall misclassification error sd   0.002865389 0.005311486
\end{verbatim}
As shown by the above results, the function prints a summary of the performance comparison between the two methods, denoted by \texttt{Classical} and \texttt{HNP}. The reported performance measures include the mean and standard deviation of the under-classification errors for classes \(i \in  [\cI - 1]\), where classes are relabeled by integers according to their importance order regardless of their original labels. We also report the corresponding violation rate for each class, estimated as the proportion of experiments in which the under-classification error exceeds the target control level. In addition, the mean and standard deviation of the overall misclassification error are reported. The target control levels and tolerances are also provided for the user to check. Furthermore, the function outputs the boxplots in Figure~\ref{fig:example3_boxplots}, which summarize the distributions of the under-classification errors and the overall classification errors. In this figure, for the under-classification errors, the black dashed line denote the corresponding control levels \(\alpha_i\), and the red dot denote the \((1-\delta_i)\)-quantiles of the associated under-classification error for \(i \in [\cI]\). To assess the effectiveness of the high-probability control, the user only needs to check whether the red dots lie  below or close to the corresponding control levels. 

From this example, we see that the violation rate of the first under-classification error (i.e., the under-classification error for class 1) is 1 for the classical method, whereas it is around \(0.22\) for the H--NP classification method, which is substantially smaller. This difference is also clearly reflected in the boxplots in Figure~\ref{fig:example3_boxplots}. Moreover, the overall classification error increases only slightly under the H--NP adjustment, from \(0.22\) to \(0.23\), indicating that the desired control is achieved at little cost. This pattern is also clearly visible in the boxplot for the overall classification error. 

We also present another example using different values of the control levels and tolerances, namely \(\alpha_1 = 0.2\), \(\alpha_2 = 0.05\), \(\delta_1 = 0.05\), and \(\delta_2 = 0.2\), while keeping all other settings unchanged. To do this, the user only needs to set \texttt{alphas <- c(0.2, 0.05)} and \texttt{deltas <- c(0.05, 0.2)} and then rerun the remaining code. The corresponding boxplots are shown in Figure~\ref{fig:example3_boxplots_1}, and the resulting performance measures for the classical and H--NP classifiers are printed as follows:
\begin{verbatim}
$classwise
                                Classical_Class_1 Classical_Class_2 H-NP_Class_1 H-NP_Class_2
control level                          0.20000000        0.05000000   0.20000000  0.050000000
tolerance                              0.05000000        0.20000000   0.05000000  0.200000000
under-classification error mean        0.23253944        0.10363368   0.17157888  0.040126520
under-classification error sd          0.03958125        0.02916384   0.01635825  0.009030163
violation rate                         0.82000000        0.98000000   0.04800000  0.142000000

$overall
                                      Classical        H-NP
overall misclassification error mean 0.22298441 0.226935227
overall misclassification error sd   0.00306058 0.004569333
\end{verbatim}
The violation rates of the H--NP classifier are below the corresponding tolerances, at the cost of only a small increase in the overall classification error. This implies that, relative to the corresponding classical base classifiers, the H--NP classifiers substantially reduce both the first and second under-classification errors with respect to the target control levels and tolerances, with little sacrifice in overall classification performance.

 In the next section, we will show that our method achieves satisfactory control of all under-classification errors under various simulation settings.

\section{Simulation studies}\label{sec:simulation}

Through the simulation examples in this section, we examine the performance of the H--NP umbrella classifiers under various settings. In particular, we investigate whether the desired high-probability control of under-classification errors is achieved and how this control affects overall classification performance. In the first simulation subsection, we consider a three-class setting and study the effects of different class-proportion ratios and training-sample splitting schemes. In the second simulation subsection, we consider a five-class setting, use the default training-sample split for the H--NP classifier, and examine its performance across different base classification methods. All results in this section are based on 
$1,000$ simulation rounds.


\subsection{Simulation 1: three-class classification}

We consider a three-class classification problem, i.e.,  $\cI = 3$, where the features follow Gaussian underlying distributions. The class labels $Y$'s are encoded as $1,2$ and $3$, in descending importance order. For each class $i \in [\cI]$, the feature vector satisfies $X\mid Y = i \sim \mathcal{N}(\mu_i, \Sigma_i)$ independently across observations. We draw $\mu_i \in \mathbb{R}^3$ uniformly from the closed unit ball
$\{u \in \mathbb{R}^3 : \|u\|_2 \le 1\}$. The covariance matrix
$\Sigma_i \in \mathbb{R}^{3 \times 3}$ is defined by $
[\Sigma_i]_{p,q} = \rho_i^{|p-q|}$, for $ p,q \in [3]$,
where $\rho_i$ is drawn uniformly from $(0,1)$. All $\mu_i$'s and
$\rho_i$'s are mutually independent.


We further consider various settings formed by combinations of the training sample size and the training-sample splitting scheme, as summarized in Table~\ref{parameter_setting}.  To evaluate classifier performance, an additional testing sample of size $10,000$ is generated for each class. We fix $(\alpha_1, \alpha_2) = (0.05, 0.1)$ and $(\delta_1, \delta_2) = (0.1, 0.1)$. In this simulation, logistic regression is used as the base learner; additional base classification methods are examined in the next simulation section. 


\begin{table}[htbp]
\centering
\small
\renewcommand{\arraystretch}{1.08}
\setlength{\tabcolsep}{5pt}
\begin{tabular}{@{} c !{\vrule width 1.2pt} c !{\vrule width 1.2pt} c c !{\vrule width 1.2pt} c c !{\vrule width 1.2pt}  c c @{}}
\Xhline{1.2pt}

\hline
 Setting & Method 
& \makecell[c]{Class 1\\$(S_{1s},S_{1t},S_{1e})$} & $N_1$
& \makecell[c]{Class 2\\$(S_{2s},S_{2t},S_{2e})$} & $N_2$
& \makecell[c]{Class 3\\$(S_{3s},S_{3t},S_{3e})$} & $N_3$ \\

\Xhline{1.2pt}

 $C_1$ & Classical & $100/-/-$ & 600 & $100/-/-$ & 600 & $100/-/-$ & 600 \\
 $C_2$ & Classical & $100/-/-$ & 300 & $100/-/-$ & 300 & $100/-/-$ & 600 \\
 
\Xhline{1.2pt}

 $T_1$ & H--NP & $50/50/-$ & 600 & $45/50/5$ & 600 & $95/-/5$ & 600 \\

 $T_2$ & H--NP & $70/30/-$ & 600 & $65/30/5$ & 600 & $95/-/5$ & 600 \\
 $T_3$ & H--NP & $30/70/-$ & 600 & $25/70/5$ & 600 & $95/-/5$ & 600 \\

 $T_4$ & H--NP & $50/50/-$ & 300 & $45/50/5$ & 300 & $95/-/5$ & 600 \\
\Xhline{1.2pt}
\end{tabular}
\caption{Settings for Simulation 1. For each class $i \in [\cI]$, $N_i$ represents the size of training sample, and the triplet $(S_{is},S_{it},S_{ie})$ provides the percentages of the score-training subset, threshold-selection subset, and error-evaluation subset, respectively.}
\label{parameter_setting}
\end{table}

The performance of the H--NP classifiers is evaluated using the average overall classification error ($R_{\rm{overall}}$), the average under-classification errors ($R_{1\star}$ and $R_{2\star}$), and the corresponding violation rates ($V_1$ and $V_2$). The results are summarized in Table~\ref{tab:sim3_results}. 
An important observation is that the
violation rates $V_1$ and $V_2$ for the H--NP umbrella classifiers are at
most, or close to, the desired tolerances $\delta_1=\delta_2=0.1$ across
settings $T_1$ to $T_4$. This indicates the effectiveness of the H--NP
umbrella algorithm in achieving high-probability control of
under-classification errors. In contrast, the classical classification
methods do not provide such probabilistic control in settings $C_1$ and
$C_2$, where the violation rates are $100\%$. This control, however,
comes with a trade-off: the H--NP classifiers generally have higher overall
classification errors, reflecting the cost of prioritizing control over the
more consequential under-classification errors.

\begin{table}[h!]
\centering

\begin{tabular}{c !{\vrule width 1.2pt} c !{\vrule width 1.2pt} c c !{\vrule width 1.2pt} c c !{\vrule width 1.2pt} c}
\Xhline{1.2pt}
Paradigm & Setting  & $R_{1\star}$ & $R_{2\star}$ & $V_1$ & $V_2$ & $R_{\text{overall}}$ \\
\Xhline{1.2pt}
\multirow{2}{*}{Classical} & $C_1$  & 0.222 & 0.344 & 1.000 & 1.000 & 0.362 \\
& $C_2$  & 0.285 & 0.709 & 1.000 & 1.000 & 0.400 \\
\Xhline{1.2pt}
\multirow{4}{*}{H-NP} & $T_1$  & 0.034 & 0.051 & 0.096 & 0.001 & 0.587 \\
& $T_2$  & 0.031 & 0.039 & 0.098 & 0.002 & 0.597 \\
& $T_3$  & 0.036 & 0.057 & 0.082 & 0.002 & 0.581 \\
& $T_4$  & 0.031 & 0.037 & 0.102 & 0.004 & 0.598 \\

\Xhline{1.2pt}
\end{tabular}
\caption{Average overall classification error $R_{\text{overall}}$, average under-classification errors $R_{1\star}$ and $R_{2\star}$, along with their corresponding violation rates $V_1$ and $V_2$ for Simulation 1.}
\label{tab:sim3_results}
\end{table}

Overall, this simulation example suggests that different class proportions and training-sample splitting schemes have no obvious effect on the performance of the H--NP classifiers. More specifically, the overall classification errors vary only minimally across settings, while effective control of all under-classification errors is maintained.

\subsection{Simulation 2: five-class classification}

We examine the adaptability of the H--NP umbrella algorithm to different base
classification methods under a five-class classification problem, i.e.,
$\cI=5$, with labels encoded as integers $1$ to $5$ in descending order of
importance. As in Simulation 1, for each class $i \in [\cI]$, observations are
independently generated from $\mathcal{N}(\mu_i,\Sigma_i)$, where
$\mu_i \in \mathbb{R}^3$ and $\Sigma_i \in \mathbb{R}^{3 \times 3}$ are
generated in the same way as in the previous simulation setting. For each
class, we independently generate a training sample of size $600$ and a testing
sample of size $10{,}000$.  We then train six types of classifiers: classical
classifiers based on logistic regression, SVM, and random forest, and H--NP
classifiers using the same three methods to construct the scoring functions. The H--NP classifiers are trained using the default training sample splitting scheme in \texttt{hnp\_umbrella()} shown in Table \ref{tab:hnp_split_default}, whereas the classical classifiers are trained on the entire training sample. The control level $\alpha_i$'s and tolerance $\delta_i$'s for all $4$ under-classification errors are set to $0.1$.




\begin{table}[h!]
\centering
\resizebox{\textwidth}{!}{%
\begin{tabular}{c !{\vrule width 1.2pt} c !{\vrule width 1.2pt}  c c c c !{\vrule width 1.2pt} c c c c !{\vrule width 1.2pt} c}
\Xhline{1.2pt}
Base method & Paradigm 

& $R_{1\star}$ & $R_{2\star}$ & $R_{3\star}$ & $R_{4\star}$ 
& $V_1$ & $V_2$ 
& $V_3$ & $V_4$ & $R_{\text{overall}}$\\
\Xhline{1.2pt}

\multirow{2}{*}{Logistic regression}
& Classical 
 & 0.491 & 0.641 & 0.113 & 0.152 
& 1.000 & 1.000 & 0.974 & 1.000 & 0.538 \\
& \textcolor{black}{H--NP}  
 & 0.065 & 0.065 & 0.015 & 0.039 
& 0.045 & 0.023 & 0.000 & 0.000 & 0.576\\

\Xhline{1.2pt}

\multirow{2}{*}{Random forest}
& Classical
 & 0.425 & 0.313 & 0.196 & 0.147 
& 1.000 & 1.000 & 1.000 & 1.000 & 0.474\\
& \textcolor{black}{H--NP}  
 & 0.065 & 0.063 & 0.049 & 0.049 
& 0.029 & 0.021 & 0.003 & 0.000 & 0.527\\

\Xhline{1.2pt}

\multirow{2}{*}{SVM}
& Classical
 & 0.307 & 0.143 & 0.142 & 0.145 
& 1.000 & 1.000 & 1.000 & 0.994 & 0.428\\
& \textcolor{black}{H--NP}  
 & 0.062 & 0.064 & 0.057 & 0.056 
& 0.038 & 0.038 & 0.002 & 0.001 & 0.487\\

\Xhline{1.2pt}

\end{tabular}
}
\caption{Average overall classification error $R_{\text{overall}}$, average under-classification errors $R_{i\star}$ for $i \in [4]$, along with their corresponding violation rates $V_i$ for $i \in [4]$ in Simulation 2.}
\label{tab:five_class_base_methods}
\end{table}

\begin{figure}[ht!]
    \centering
    \begin{subfigure}[b]{0.75\textwidth}
    \includegraphics[width=1\textwidth]{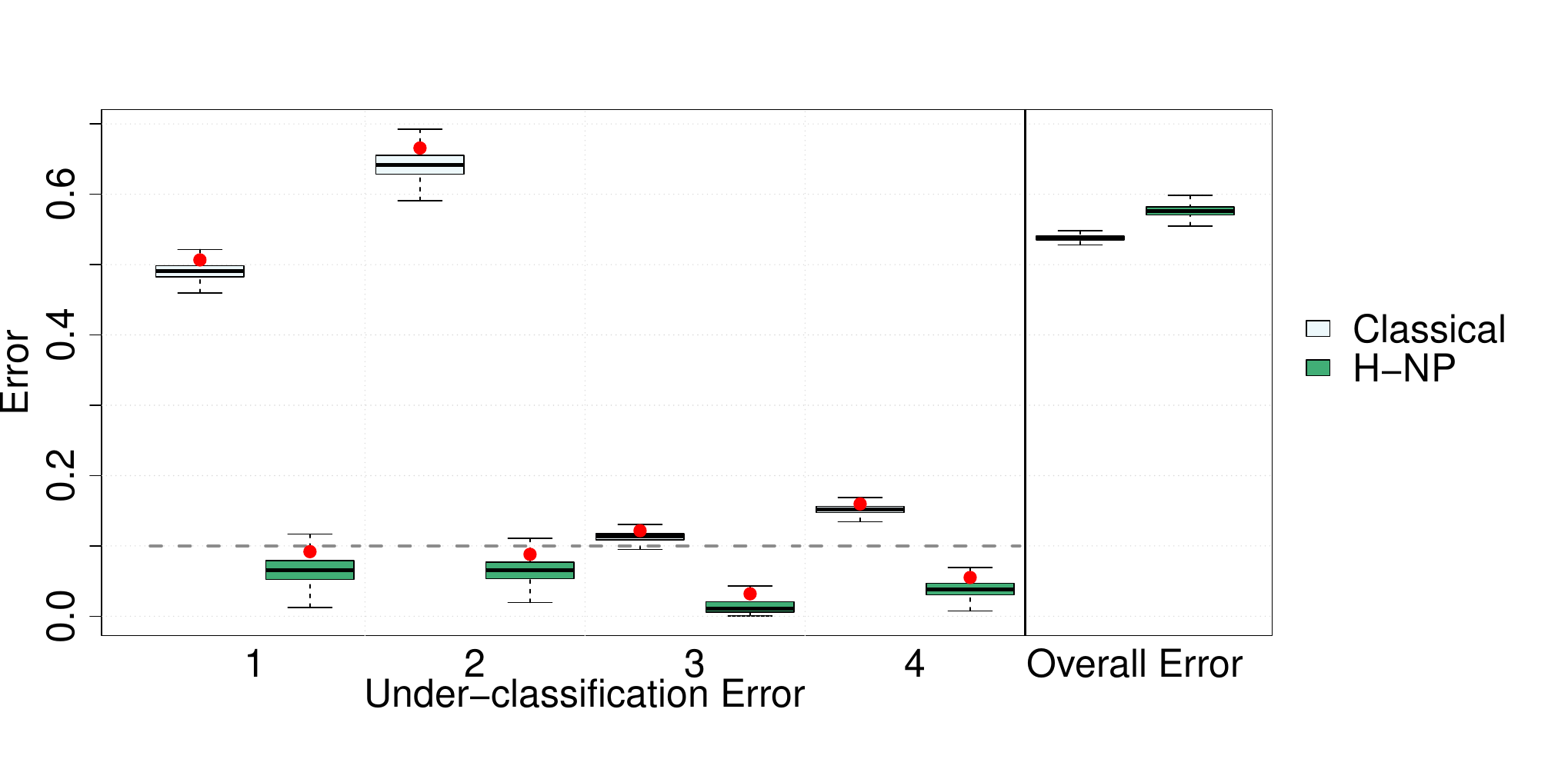}
    \caption{Logistic regression}
    \end{subfigure}
    \begin{subfigure}[b]{0.75\textwidth}
    \includegraphics[width=1\textwidth]{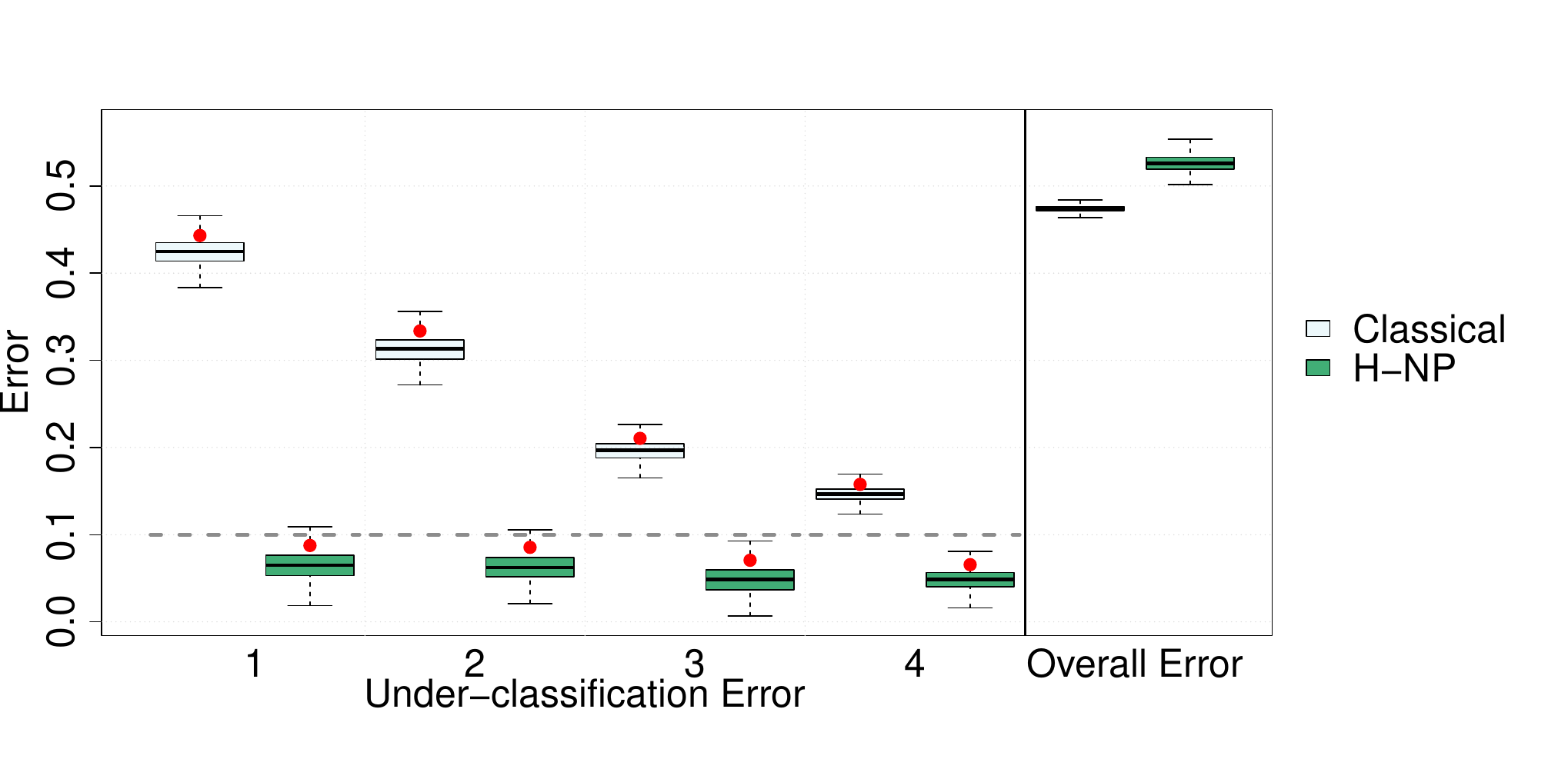}
    \caption{Random forest}
    \end{subfigure}
    \begin{subfigure}[b]{0.75\textwidth}
    \includegraphics[width=1\textwidth]{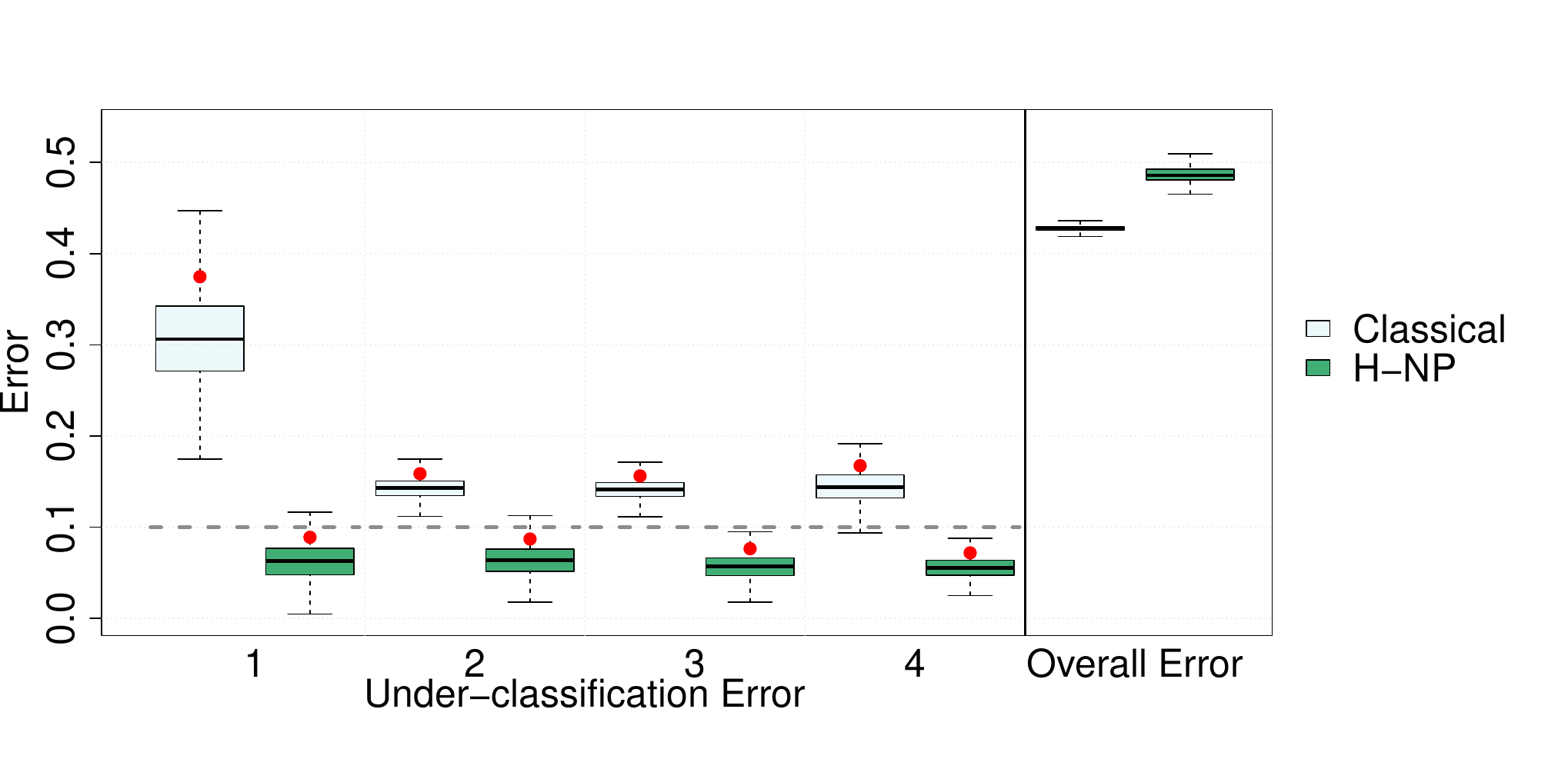}
    \caption{SVM}
    \end{subfigure}
    \caption{\footnotesize Distributions of the approximate errors for classifiers using different base learners in Simulation 2.}
    \label{fig:five_methods}
\end{figure}

Table~\ref{tab:five_class_base_methods} summarizes the distributions of the
errors of interest, while Figure~\ref{fig:five_methods} presents the detailed
distributions using boxplots. All four under-classification errors of the three H--NP classifiers are well controlled below the predetermined  control levels with violation rates less than tolerances. This is further confirmed in Figure~\ref{fig:five_methods}: in every panel, the $1 - \delta_i$ quantiles (red dots) of the under-classification errors lie below $\alpha_i$ (dashed line). In contrast, the classical classifiers do not
provide control over these errors, regardless of the method used. Meanwhile,
the overall classification errors of the H--NP classifiers are larger than
those of their classical counterparts, as expected, due to the trade-off
between prioritized error control and overall classification performance. Furthermore, comparing the average overall classification error
$R_{\rm overall}$ in Table~\ref{tab:five_class_base_methods} shows that when a classical classifier achieves better
overall performance, the corresponding
H--NP classifier using the same method as the base learner also tends to
perform better among the H--NP classifiers.

In conclusion, the choice of base classification method has no obvious effect
on the effectiveness of controlling all under-classification errors. Meanwhile,
the overall performance of the H--NP classifiers with different base learners
varies consistently with the performance of their corresponding classical
classifiers.

\section{Real data applications}\label{sec:real
_data}

This section illustrates the practical use of the H--NP umbrella algorithm through two real-data applications involving ordered multi-class decision problems. The first application considers diabetes status prediction using the Diabetes Health Indicators Dataset, where pre-diabetes, diabetes, and healthy status are ordered according to their intervention priority. The second application analyzes the South German Credit dataset by constructing a five-class ordered outcome that combines credit-risk status and loan amount, with the goal of limiting the risk of approving loans for bad-risk applicants or assigning loan amounts above a customer's safe loan-amount band. Overall, the results highlight the ability of the H--NP framework to prioritize consequential error types while maintaining flexibility across different base classification methods.

\subsection{Diabetes status prediction}

Under-classification control is particularly important in medical applications. In this case study, we apply the H--NP umbrella algorithm to construct classifiers for identifying stages of diabetes progression. Predictive models \citep{xie2019building} have the potential to facilitate early diagnosis and intervention, thereby ultimately reducing healthcare costs. We use the Diabetes Health Indicators Dataset on Kaggle \citep{CDC2015}, which contains \(253{,}680\) survey responses from the U.S. Centers for Disease Control and Prevention's (CDC) 2015 Behavioral Risk Factor Surveillance System (BRFSS). The features include health-related indicators, lifestyle risk factors, and clinical variables, such as blood pressure, smoking status, history of stroke, heart disease, and demographic characteristics, which are used to predict three labels: pre-diabetes, diabetes, and healthy. The proportions of these three classes in the sample are \(1.83\%\), \(13.93\%\), and \(84.24\%\), respectively.

For intervention purposes, we treat pre-diabetes as the most important class, since it is the critical stage at which preventive action may help avoid progression to diabetes and is often more difficult to detect. The next most important class is diabetes, for which timely medical care is needed, followed by the healthy class. In this setting, missed diagnoses may lead to delayed intervention, and our H--NP classification methods can help facilitate early diagnosis and timely intervention by providing guarantees on under-classification errors. Accordingly, we define class 1 as pre-diabetes, class 2 as diabetes, and class 3 as healthy. This analysis sets \(\alpha_1 = 0.4\), \(\alpha_2 = 0.2\), and \(\delta_1 = \delta_2 = 0.2\).

We repeatedly split the data at random into training and testing sets with a ratio of \(0.05/0.95\), and all experiments are repeated over 100 runs. Here, we aim to illustrate the performance of our method under a relatively small training sample size; hence, this train--test ratio is adopted. For the classical classifiers used for comparison, all training samples are used directly for model fitting without additional splitting.

To construct the H--NP classifier with logistic regression and SVM as the base methods, we use the default settings in \texttt{hnp\_umbrella()}, as discussed in Section~\ref{sec:hnp_package}, including the sample-splitting scheme for score training, threshold selection, and evaluation. For the random forest base method, however, the default score-training procedure in \texttt{hnp\_umbrella()} does not perform well, as it fails to provide good separation between classes; consequently, the resulting H--NP classifier also performs poorly. Therefore, following the strategy discussed in Example~2 of Section~\ref{sec:implementation_details}, we use a pre-trained scoring function based on random forest instead of the default score-training procedure built into \texttt{hnp\_umbrella()}. Specifically, for score training, we use \(50\%\), \(45\%\), and \(95\%\) of the training data from classes 1, 2, and 3, respectively. The resulting trained scoring function, together with the remaining training data, is then supplied to \texttt{hnp\_umbrella()} to construct the H--NP classifier. Under this setting, the subsequent sample splitting for threshold selection and evaluation follows the default configuration described in Table~\ref{tab:hnp_split_default}.

\begin{figure}[!ht]
\centering
\hfill
\begin{subfigure}[b]{0.48\textwidth}
    \centering
    \includegraphics[width=\textwidth]{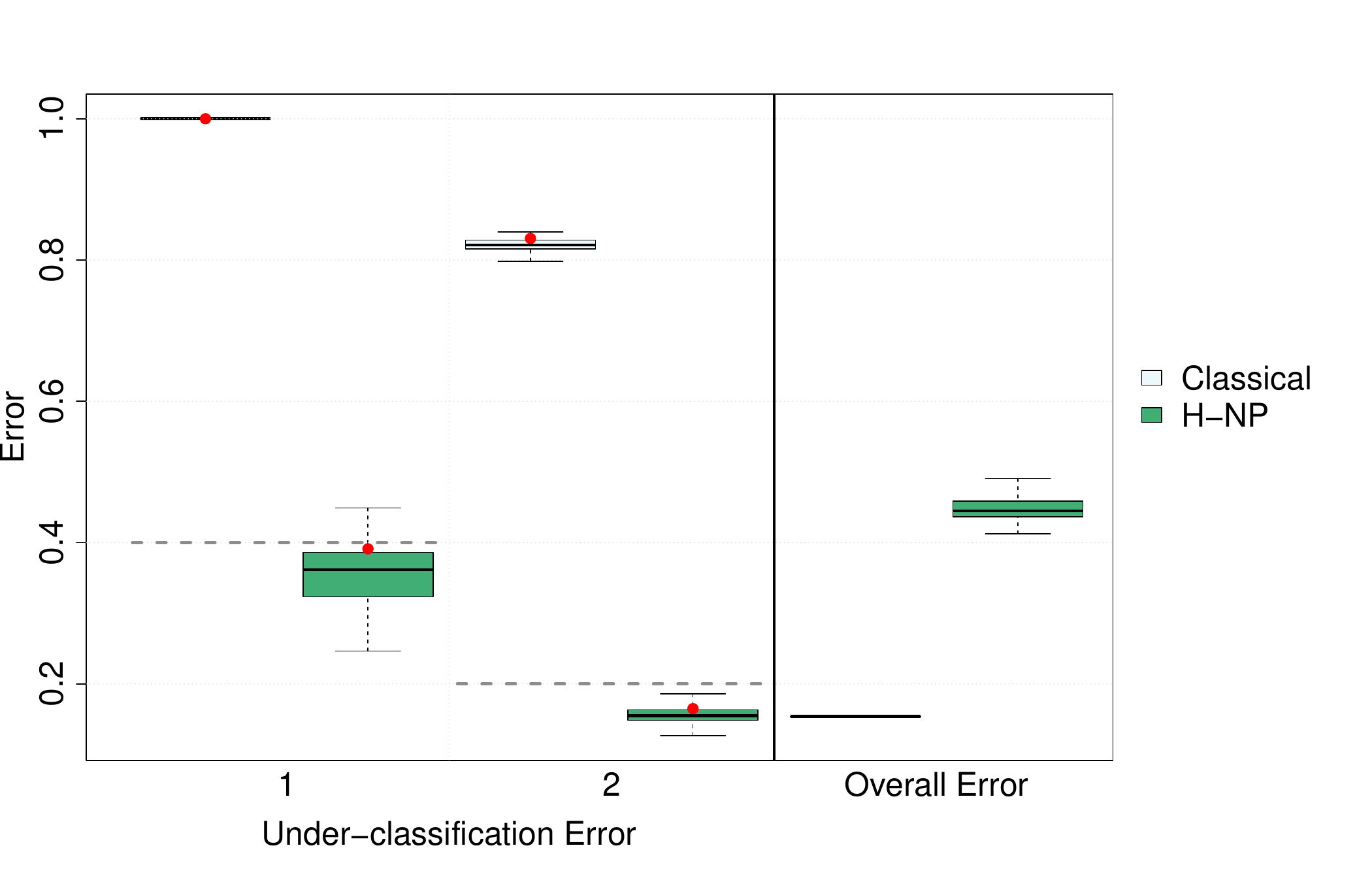}
    \caption{Logistic regression}
    \label{fig:diabetes-rf}
\end{subfigure}
\hfill
\begin{subfigure}[b]{0.48\textwidth}
    \centering
    \includegraphics[width=\textwidth]{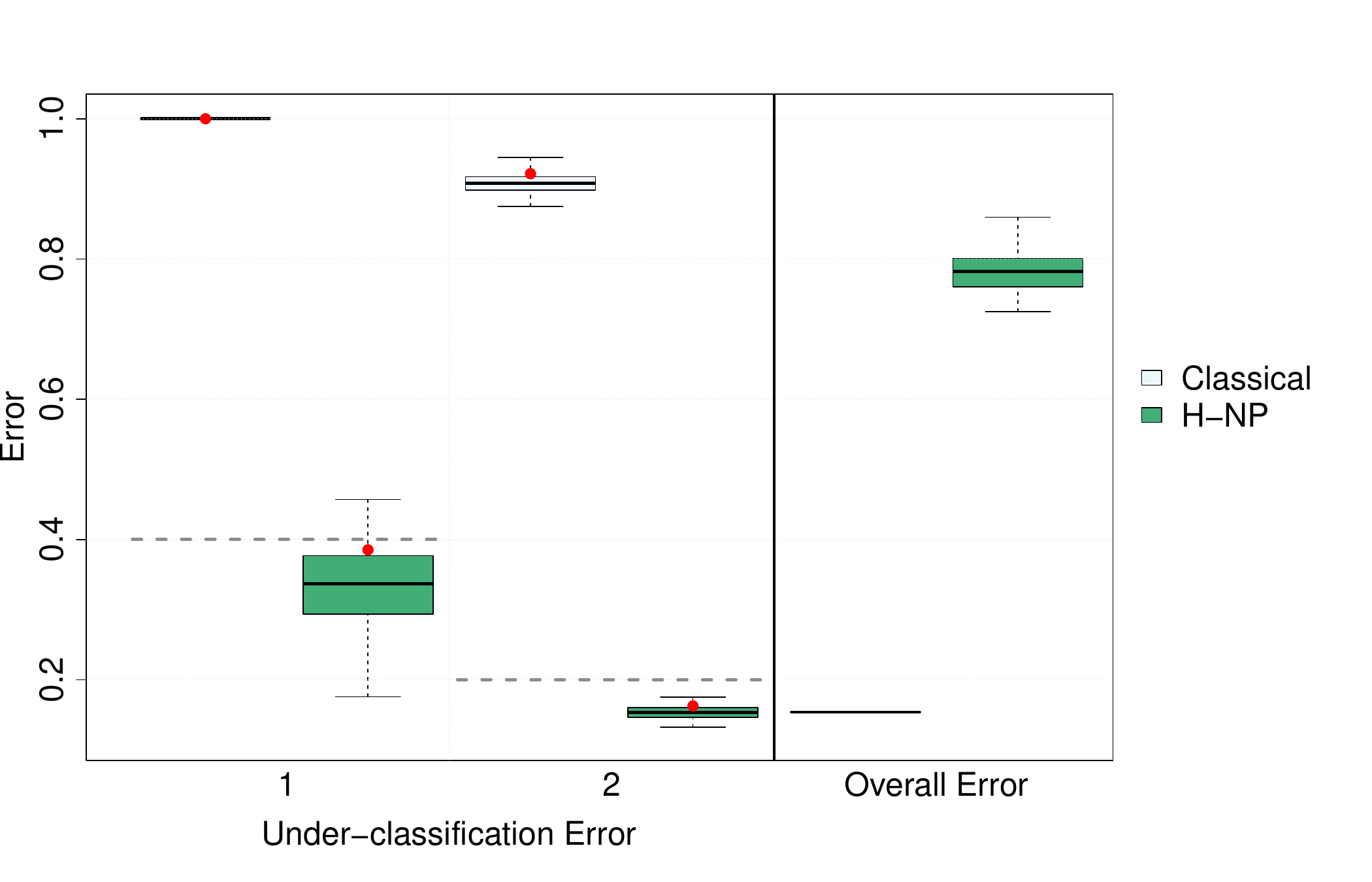}
    \caption{SVM}
    \label{fig:diabetes-svm}
\end{subfigure}
\hfill
\begin{subfigure}[b]{0.48\textwidth}
    \centering
    \includegraphics[width=\textwidth]{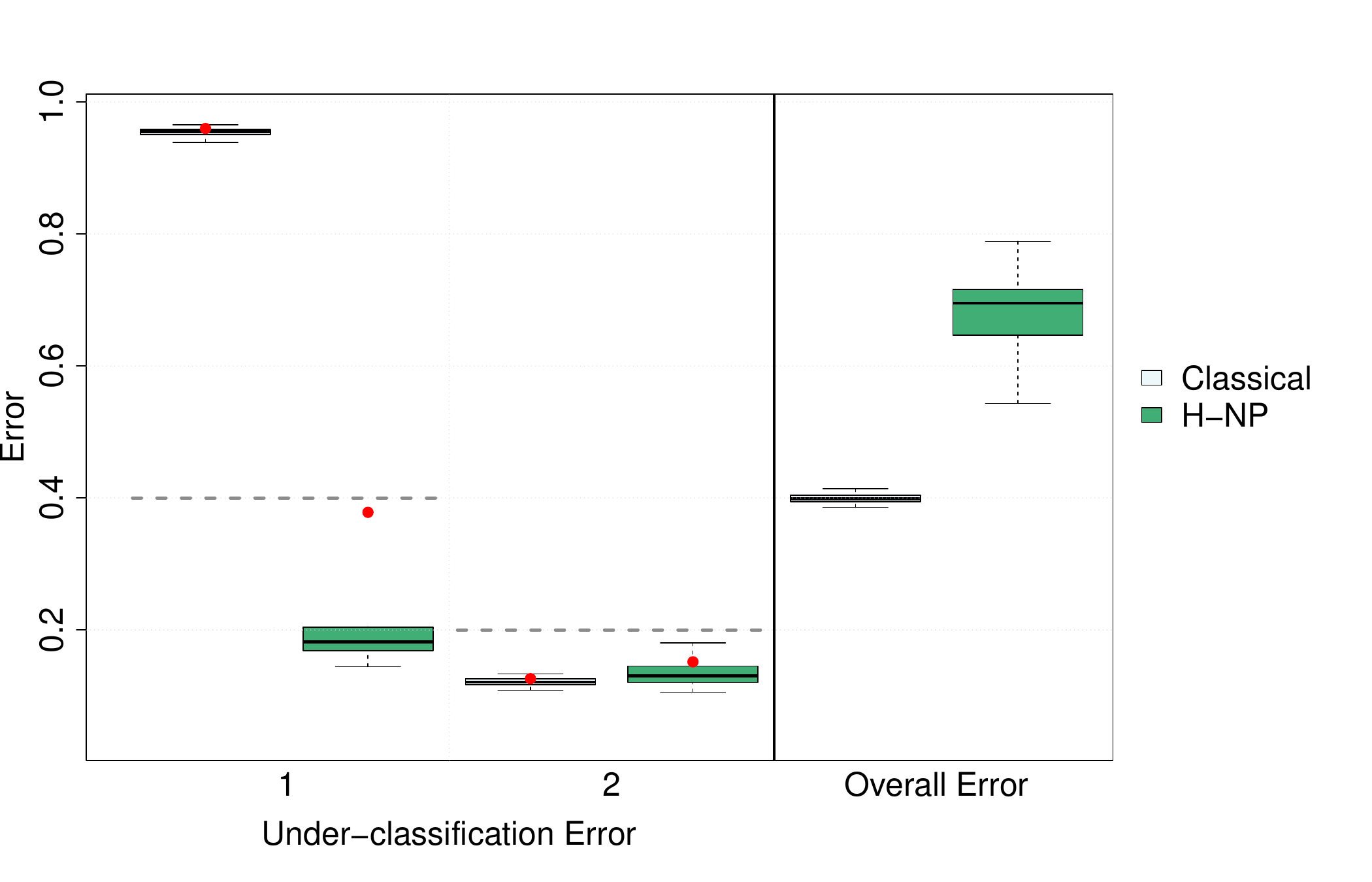}
    \caption{Random forest (pre-trained)}
    \label{fig:diabetes-rf}
\end{subfigure}
\caption{\footnotesize Distribution of the approximate errors for diabetes status prediction under different base classification methods.}
\label{fig:diabetes_compare}
\end{figure}

Figure~\ref{fig:diabetes_compare} presents boxplots of the distributions of the under-classification errors and the overall classification errors, comparing the H--NP classifiers with the corresponding classical classifiers.  Table~\ref{tab:diabetes_base_methods} further summarizes the average errors and violation rates. The results show that all classical classifiers produce an under-classification error close to 1 for class 1. For class 2, the under-classification error is close to $0.8$ for the classical classifiers based on logistic regression and SVM, but not for the random forest classifier. By contrast, for the H--NP classifiers, the red dots in Figure~\ref{fig:diabetes_compare}, which denote the \(1-\delta_i\) quantiles (i.e., the \(80\%\) quantiles), all lie at or below the corresponding target control levels indicated by the black dashed lines for all under-classification errors across all base classification methods. Correspondingly, all violation rates reported in Table~\ref{tab:diabetes_base_methods} are also below the target levels. These results illustrate that the H--NP classifiers are able to provide high-probability control of the under-classification errors, a property that is especially important for enabling timely and effective intervention in diabetes progression.

\begin{table}[h!]
\centering

\begin{tabular}{c !{\vrule width 1.2pt} c !{\vrule width 1.2pt}  c c !{\vrule width 1.2pt} c c !{\vrule width 1.2pt} c}
\Xhline{1.2pt}
Base method & Paradigm 

& $R_{1\star}$ & $R_{2\star}$
& $V_1$ & $V_2$ & $R_{\text{overall}}$\\
\Xhline{1.2pt}

\multirow{2}{*}{Logistic regression}
& Classical 
 & 1.000 & 0.822
& 1.000 & 1.000 & 0.154 \\
& \textcolor{black}{H--NP}  
 & 0.357 & 0.155
& 0.130 & 0.000 & 0.448\\

\Xhline{1.2pt}

\multirow{2}{*}{SVM}
& Classical 
 & 1.000 & 0.909
& 1.000 & 1.000 & 0.154 \\
& \textcolor{black}{H--NP}  
 & 0.330 & 0.154
& 0.110 & 0.000 & 0.783\\

\Xhline{1.2pt}

\multirow{2}{*}{Random forest (pre-trained)}
& Classical 
 & 0.954 & 0.121
& 1.000 & 0.000 & 0.400 \\
& \textcolor{black}{H--NP}  
 & 0.228 & 0.124
& 0.080 & 0.000 & 0.667\\

\Xhline{1.2pt}

\end{tabular}

\caption{Average overall classification error $R_{\text{overall}}$, average under-classification errors $R_{i\star}$'s, along with their corresponding violation rates $V_i$'s for the Diabetes status prediction.}
\label{tab:diabetes_base_methods}
\end{table}

\subsection{South German Credit Data Analysis}

We aim to demonstrate the use of the H--NP Umbrella Algorithm when the number of classes satisfies $\cI > 3$, using the South German Credit dataset \citep{statlog144}. The South German Credit dataset is a credit-scoring dataset commonly used to study whether a loan applicant is likely to have good or bad credit risk. In this case study, we focus on classifying each applicant into a safe loan-amount band. The dataset contains $1,000$ observations and $21$ variables, including credit risk, loan amount, loan duration, loan purpose, number of existing credits, savings-account status, and checking-account status. Among the $1,000$ applicants, 700 are labeled as good credit risks and 300 are labeled as bad credit risks.

To address our goal, we construct a five-class outcome using both the binary credit-risk label and the credit amount, where credit amount refers to the size of the loan. The most important class, class $1$, consists of applicants whose credit risk is labeled as bad; these applicants are treated as cases for which the loan should not be approved. Among applicants with good credit risk, we further divide the observations into four classes according to the credit amount, so that a reasonable loan-amount band can be assigned. Specifically, for applicants labeled as good credit risks, we split the credit amount into four ordinal categories using sample quantiles. The cut points are placed at the 25th, 50th, and 75th percentiles of the credit-amount distribution among good-risk applicants. These categories form classes $2 - 5$, where class $2$ represents the smallest loan-amount band and class $5$ represents the largest loan-amount band. We note that the credit amount used in this analysis is obtained after a monotone transformation and therefore does not represent the original loan amount directly. For this reason, a quantile-based discretization is used to define the ordinal loan-amount categories.

\begin{figure}
 \centering
 \begin{minipage}{1\linewidth}
 \centering
		\includegraphics[width=10cm]{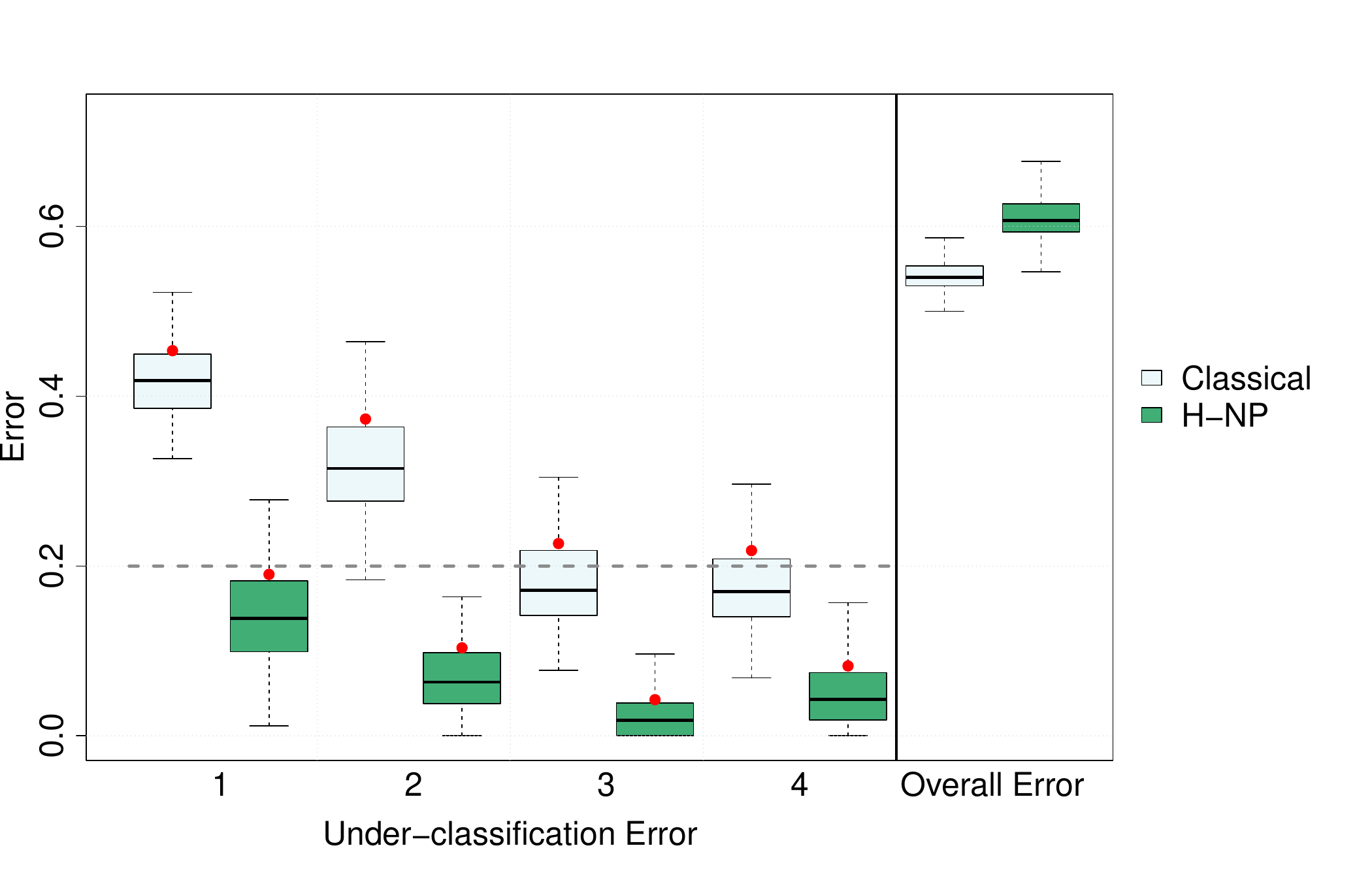}
	\end{minipage}

	\begin{minipage}{1\linewidth}
		\centering
\footnotesize
\begin{tabular}{ c !{\vrule width 1.2pt}  c c c c !{\vrule width 1.2pt} c c c c !{\vrule width 1.2pt} c}
\Xhline{1.2pt}
 Paradigm 

& $R_{1\star}$ & $R_{2\star}$ & $R_{3\star}$ & $R_{4\star}$
& $V_1$ & $V_2$ & $V_3$ & $V_4$ & $R_{\text{overall}}$\\
\Xhline{1.2pt}

 Classical 
 & 0.419 & 0.318 & 0.180 & 0.177 
& 1.000 & 0.960 & 0.340 & 0.300 & 0.541 \\
 \textcolor{black}{H--NP}  
 & 0.144 & 0.068 & 0.025 & 0.052 
& 0.120 & 0.000 & 0.000 & 0.000 & 0.611\\

\Xhline{1.2pt}

\end{tabular}

	\end{minipage}\hfill
	\caption{ \footnotesize Approximate error distributions, average errors, and violation rates for the South German Credit data analysis using logistic regression as the base classification method.}\label{fig:german_compare_logistic}
 \end{figure}

Under this construction, the resulting class labels represent increasingly larger loan-amount bands that may be safely approved for bank customers, while class 
$1$ represents applicants with bad credit risk, for whom loan approval is not recommended.  By controlling the first under-classification error, the bank can control the risk of approving loans for potentially high-risk applicants. Similarly, by controlling the 
$i$-th under-classification error, the bank can control the risk of approving a loan amount that exceeds the applicant’s safe loan-amount band, thereby reducing the likelihood of unaffordable lending. This analysis sets \(\alpha_i =  0.2\), and \(\delta_i = 0.2\) for all $i \in [\cI - 1]$.  We randomly split the data into training and testing sets using a \(0.70/0.30\) ratio, and repeat all experiments over 100 independent runs. All other settings are kept the same as in the previous real-data case study on diabetes status prediction. Using logistic regression as the base classification method, Figure~\ref{fig:german_compare_logistic} presents the distributions of the under-classification errors and the overall classification errors for the H--NP classifiers and compares them with those of the corresponding classical logistic regression classifier. The figure also summarizes the average errors and violation rates. We observe that the \(80\%\) quantiles of the under-classification errors all lie below their corresponding control levels, and that the violation rates are uniformly smaller than the desired tolerances. Additional results using random forest and SVM as the base methods are presented in Supplementary Figure~\ref{fig:german_compare} and Table~\ref{tab:german_credit_base_methods}. Likewise, none of these methods has violation rates exceeding the corresponding tolerance levels. These results indicate that the H--NP classifiers provide high-probability control of the under-classification errors, thereby supporting safer loan-approval decisions across different loan-amount bands.

\section{Summary}

This paper introduces \href{https://cran.r-project.org/web/packages/HNPclassifier/index.html}{\pkg{HNPclassifier}}, an R package for hierarchical Neyman--Pearson classification. The package implements the H--NP umbrella algorithm, which constructs classifiers that prioritize the control of under-classification errors. The main function, \code{hnp\_umbrella()}, allows users to construct H--NP classifiers using several built-in base learners. The package also supports user-supplied pretrained models, scoring functions, and score matrices, making it flexible enough to incorporate customized or externally trained learning algorithms. In addition, \code{hnp\_summary()} provides performance summaries, including confusion matrices, under-classification errors, remaining errors, and overall classification errors, while \code{hnp\_boxplot()} visualizes the empirical distributions of these errors across repeated experiments.

Simulation studies confirm the effectiveness of the proposed implementation. These results also illustrate the trade-off between minimizing overall classification error and controlling priority-specific under-classification errors. Furthermore, the real-data applications illustrate the practical relevance of the package. In the diabetes status prediction example, H--NP classifiers substantially reduce the risk of under-classifying clinically important cases, thereby supporting earlier diagnosis and intervention. In the South German Credit data analysis, the method is applied to a five-class safe loan-amount classification problem, where H--NP classifiers provide high-probability control of under-classification errors across loan-amount bands. These examples demonstrate that \href{https://cran.r-project.org/web/packages/HNPclassifier/index.html}{\pkg{HNPclassifier}} is applicable to diverse domains in which ordered class labels and asymmetric error priorities arise naturally.

Overall, \href{https://cran.r-project.org/web/packages/HNPclassifier/index.html}{\pkg{HNPclassifier}} provides a practical and extensible toolkit for multi-class classification with high-probability control of under-classification errors. By combining the flexibility of widely used classification methods with the error-control guarantees of the H--NP framework, the package enables users to construct classifiers that better reflect the priorities and risk structures present in real-world decision-making problems.

\bibliography{RJreferences}
\bibliographystyle{apalike}

\clearpage

\appendix
\appendixpage

\section{Additional real data results}

\begin{figure}[!ht]
\centering
\begin{subfigure}[b]{0.48\textwidth}
    \centering
    \includegraphics[width=\textwidth]{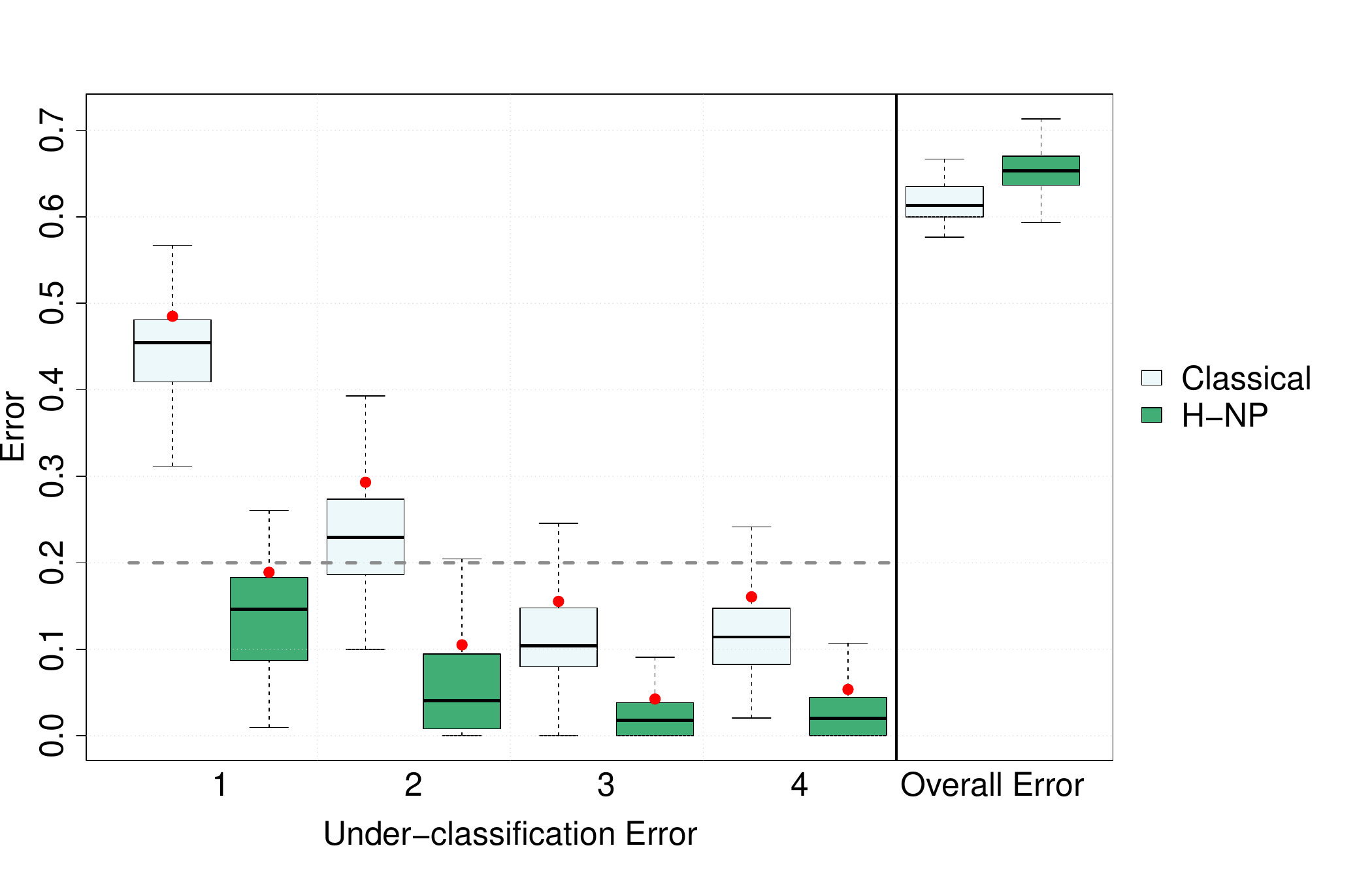}
    \caption{Random forest}
    \label{fig:german-rf}
\end{subfigure}
\hfill
\begin{subfigure}[b]{0.48\textwidth}
    \centering
    \includegraphics[width=\textwidth]{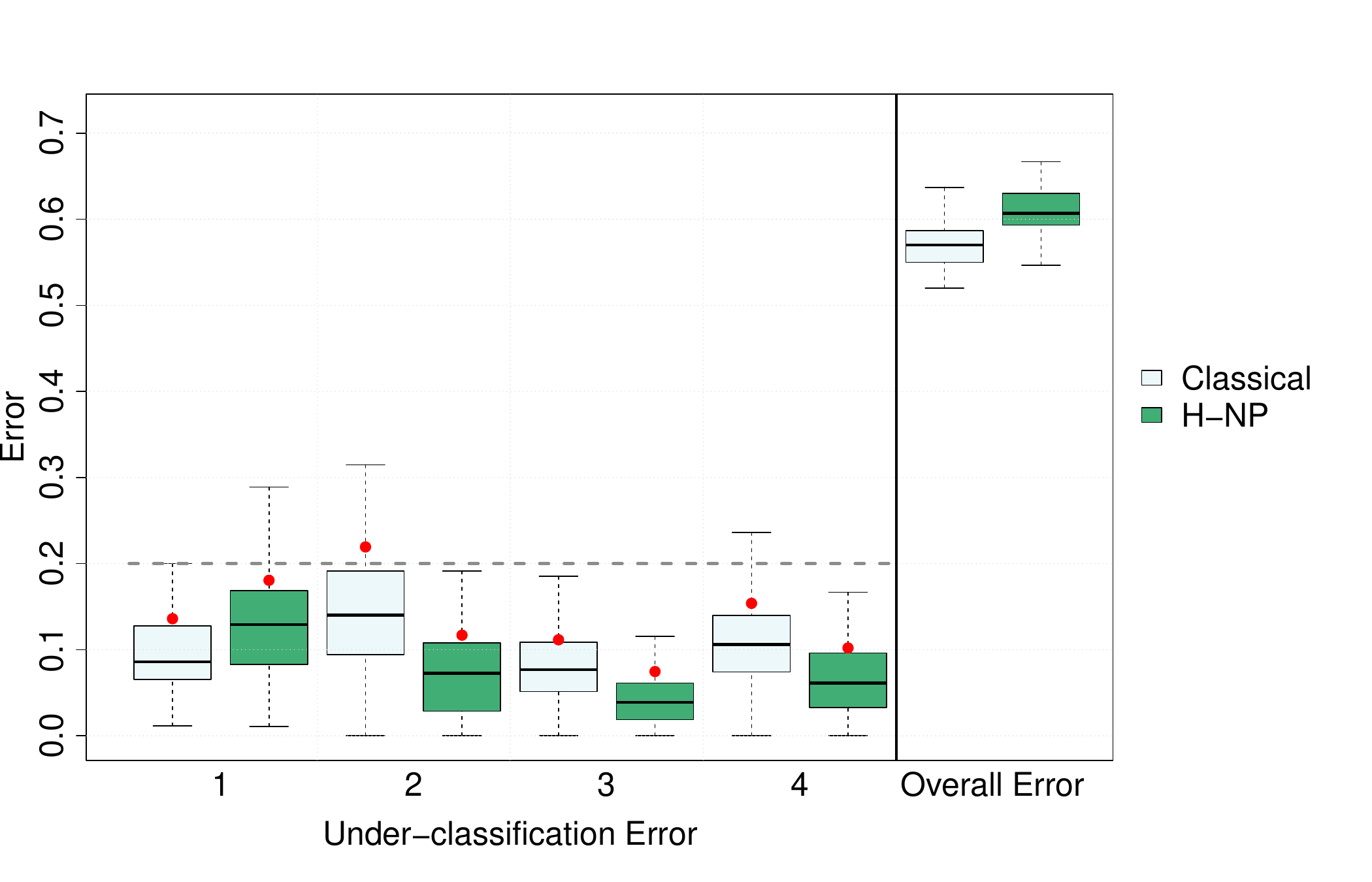}
    \caption{SVM}
    \label{fig:german-svm}
\end{subfigure}

\caption{\footnotesize Distribution of the approximate errors for South German credit data analysis under different base classification methods.}
\label{fig:german_compare}
\end{figure}

\begin{table}[h!]
\centering
\resizebox{\textwidth}{!}{%
\begin{tabular}{c !{\vrule width 1.2pt} c !{\vrule width 1.2pt}  c c c c !{\vrule width 1.2pt} c c c c !{\vrule width 1.2pt} c}
\Xhline{1.2pt}
Base method & Paradigm 

& $R_{1\star}$ & $R_{2\star}$ & $R_{3\star}$ & $R_{4\star}$
& $V_1$ & $V_2$ & $V_3$ & $V_4$ & $R_{\text{overall}}$\\
\Xhline{1.2pt}

\multirow{2}{*}{Random forest}
& Classical 
 & 0.094 & 0.146 & 0.082 & 0.113 
& 0.000 & 0.230 & 0.010 & 0.070 & 0.573 \\
& \textcolor{black}{H--NP}  
 & 0.128 & 0.075 & 0.045 & 0.064 
& 0.100 & 0.020 & 0.010 & 0.000 & 0.610\\

\Xhline{1.2pt}

\multirow{2}{*}{SVM}
& Classical 
 & 0.449 & 0.234 & 0.113 & 0.119 
& 1.000 & 0.680 & 0.070 & 0.030 & 0.618 \\
& \textcolor{black}{H--NP}  
 & 0.140 & 0.061 & 0.025 & 0.028 
& 0.150 & 0.020 & 0.000 & 0.000 & 0.654\\

\Xhline{1.2pt}

\end{tabular}
}
\caption{Average overall classification error $R_{\text{overall}}$, average under-classification errors $R_{i\star}$'s, along with their corresponding violation rates $V_i$'s for the German credit data analysis.}
\label{tab:german_credit_base_methods}
\end{table}

\section{Additional simulation results}\label{sec:extra-sim}

\begin{figure}[h!]
\centering

\begin{subfigure}{0.95\textwidth}
    \centering
    \includegraphics[width=\textwidth]{\detokenize{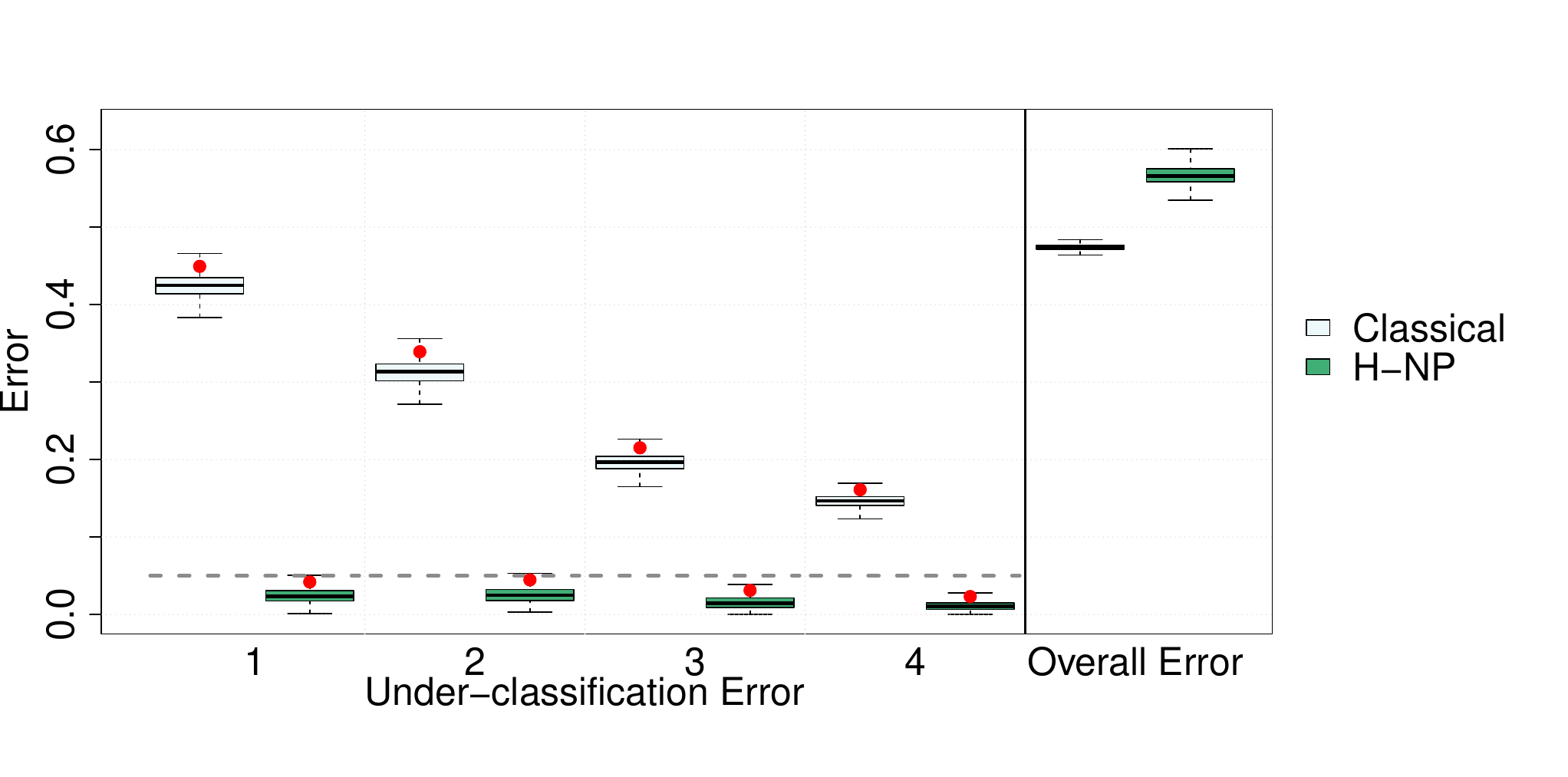}}
    \caption{Random forest}
\end{subfigure}

\vspace{0.5em}

\begin{subfigure}{0.95\textwidth}
    \centering
    \includegraphics[width=\textwidth]{\detokenize{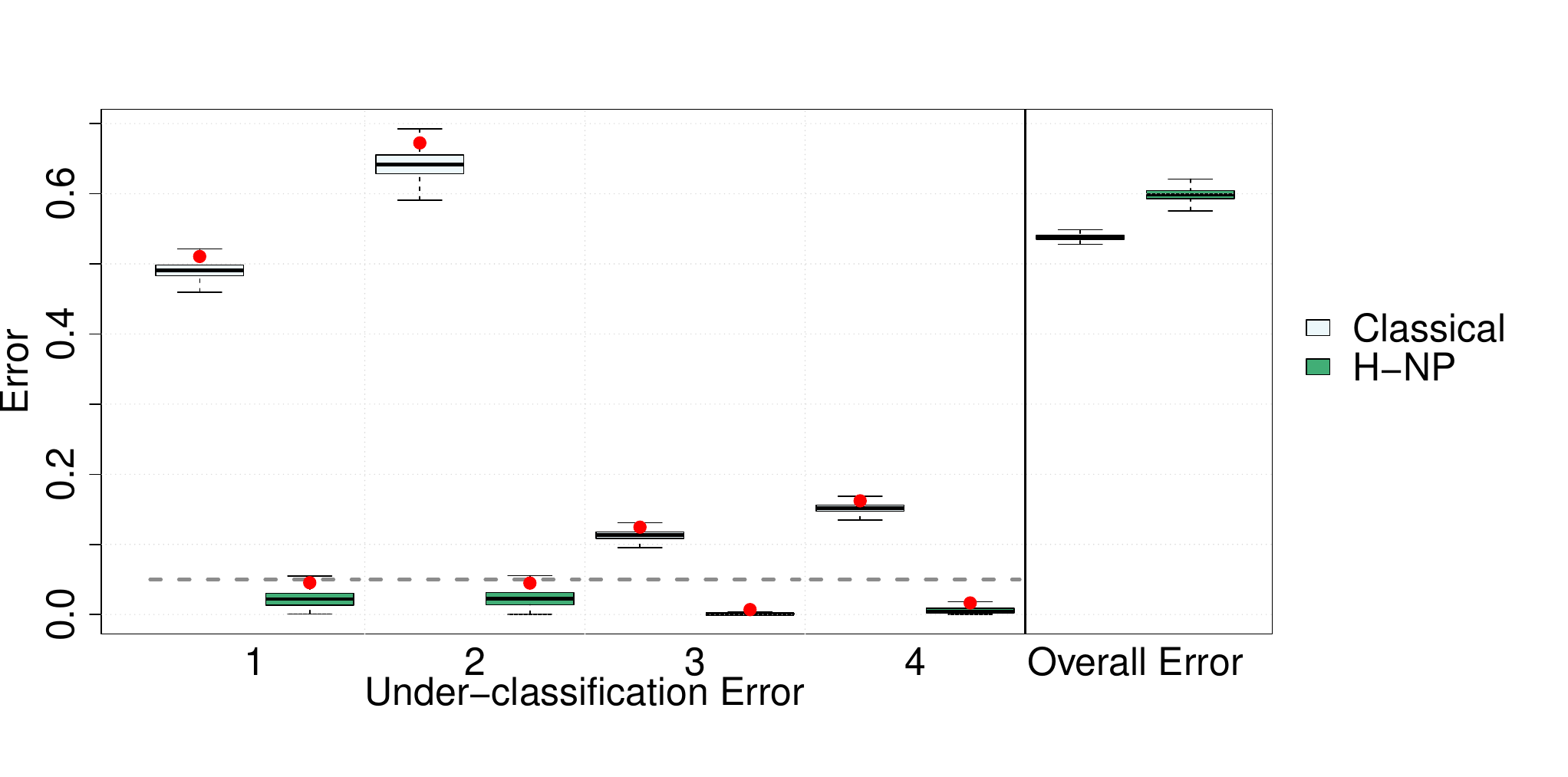}}
    \caption{Logistic regression}
\end{subfigure}

\vspace{0.5em}

\begin{subfigure}{0.95\textwidth}
    \centering
    \includegraphics[width=\textwidth]{\detokenize{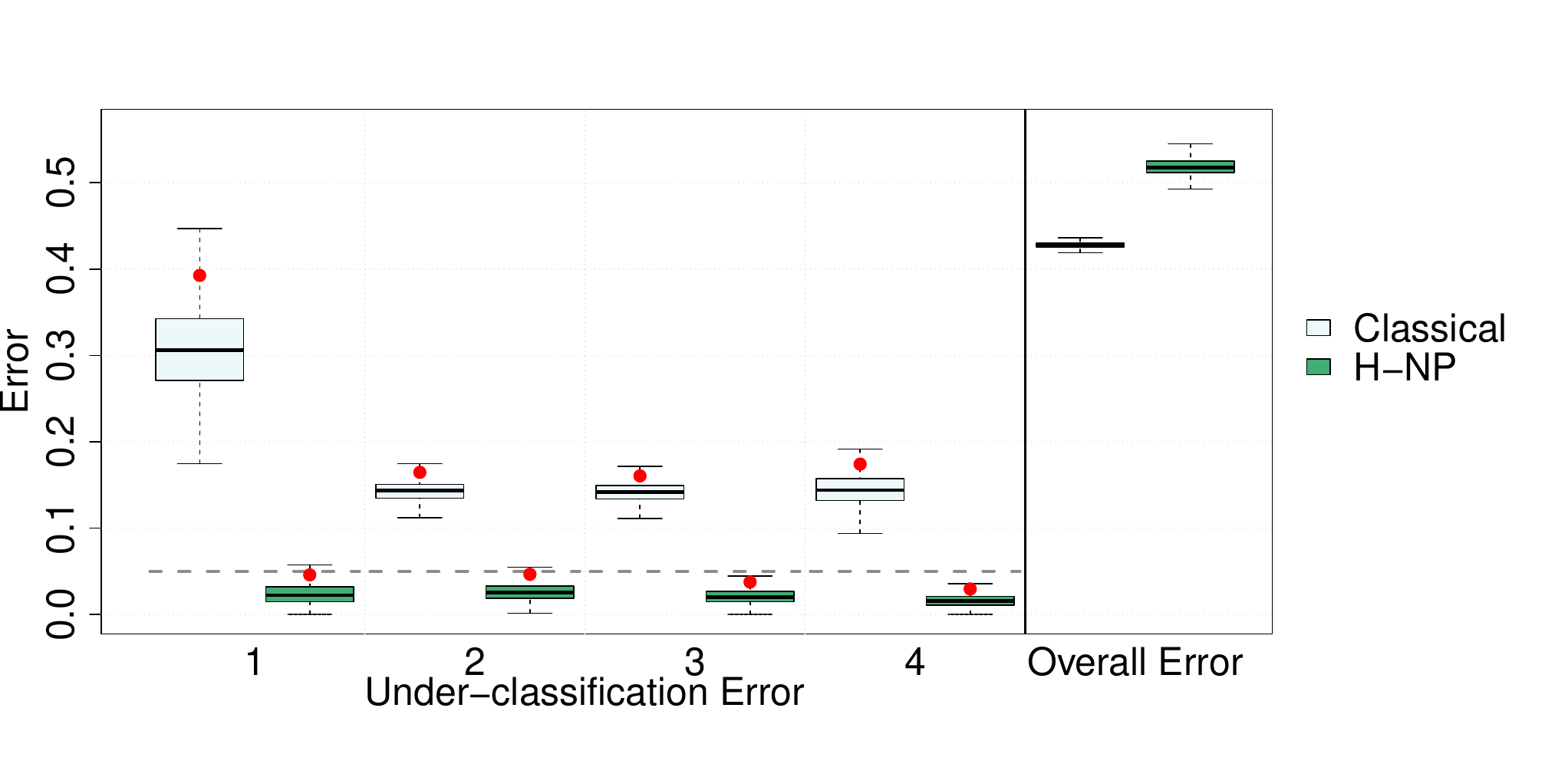}}
    \caption{SVM}
\end{subfigure}

\caption{Boxplots of under-classification errors and overall classification errors for different base methods in Section \ref{sec:extra-sim}, with $\alpha_i=\delta_i=0.05$ for all $i\in[4]$.}
\label{fig:five_class_base_methods_boxplots}
\end{figure}

In this section, we present another simulation setting for 5-class classification, similar to the one used in Simulation 2. The only modification is that the control level $\alpha_i$'s and tolerance $\delta_i$'s are all set to $0.05$. The results are summarized in Table \ref{tab:five_class_base_methods_005} and Figure \ref{fig:five_class_base_methods_boxplots}. This simulation again confirms the high-probability control of under-classification errors by H--NP classifiers, while the increase in overall classification error remains mild.

\begin{table}[h!]
\centering
\resizebox{\textwidth}{!}{%
\begin{tabular}{c !{\vrule width 1.2pt} c !{\vrule width 1.2pt}  c c c c !{\vrule width 1.2pt} c c c c !{\vrule width 1.2pt} c}
\Xhline{1.2pt}
Base method & Paradigm 
& $R_{1\star}$ & $R_{2\star}$ & $R_{3\star}$ & $R_{4\star}$ 
& $V_1$ & $V_2$ 
& $V_3$ & $V_4$ & $R_{\text{overall}}$\\
\Xhline{1.2pt}

\multirow{2}{*}{Logistic regression}
& Classical 
& 0.491 & 0.641 & 0.113 & 0.152
& 1.000 & 1.000 & 1.000 & 1.000 & 0.538 \\
& \textcolor{black}{H--NP}  
& 0.023 & 0.023 & 0.002 & 0.006
& 0.028 & 0.017 & 0.000 & 0.000 & 0.598 \\

\Xhline{1.2pt}

\multirow{2}{*}{Random forest}
& Classical
& 0.425 & 0.313 & 0.196 & 0.147
& 1.000 & 1.000 & 1.000 & 1.000 & 0.474 \\
& \textcolor{black}{H--NP}  
& 0.025 & 0.026 & 0.016 & 0.011
& 0.015 & 0.025 & 0.001 & 0.000 & 0.567 \\

\Xhline{1.2pt}

\multirow{2}{*}{SVM}
& Classical
& 0.307 & 0.143 & 0.142 & 0.145
& 1.000 & 1.000 & 1.000 & 1.000 & 0.428 \\
& \textcolor{black}{H--NP}  
& 0.024 & 0.027 & 0.021 & 0.016
& 0.031 & 0.032 & 0.002 & 0.000 & 0.519 \\

\Xhline{1.2pt}
\end{tabular}
}
\caption{Average overall classification error $R_{\text{overall}}$, average under-classification errors $R_{i\star}$, and violation rates $V_i$ for $i \in [4]$ in Section \ref{sec:extra-sim}, with $\alpha_i = \delta_i = 0.05$ for all $i \in [4]$.}
\label{tab:five_class_base_methods_005}
\end{table}

\end{document}